\documentclass[12pt]{article}
\pdfoutput=1
\usepackage{jheppub}
\usepackage{amsmath,amssymb}

\makeatletter

\@addtoreset{equation}{section}
\makeatother

\newcommand{\beq}{\begin{equation}}
\newcommand{\eeq}{\end{equation}}

\begin{document}

\baselineskip=18pt  
\baselineskip 0.7cm

\begin{titlepage}

\setcounter{page}{0}

\renewcommand{\thefootnote}{\fnsymbol{footnote}}

\begin{flushright}
\end{flushright}

\vskip 1.5cm

\begin{center}
{\LARGE \bf
Supersymmetric Partition Functions

 \vskip 0.5cm
and a
\vskip 0.5 cm
String Theory in 4 Dimensions}

\vskip 1.5cm 

{\large
Cumrun Vafa
\\
\medskip
}

\vskip 0.5cm

{
\it
Jefferson Physical Laboratory, Harvard University, Cambridge, MA 02138, USA\\
\medskip
}

\end{center}

\centerline{{\bf Abstract}}
\medskip
\noindent
We propose a novel string theory propagating in a non-commutative deformation of the four dimensional space 
$T^*{\bf T}^2$ whose scattering states correspond to superconformal theories in 5 dimensions and the
scattering amplitudes compute superconformal indices of the corresponding 5d theories. 
The superconformal theories are obtained by M-theory
compactifications on singular CY 3-folds or equivalently from a web of 5-branes of type IIB strings.  The cubic interaction of this string theory for primitive winding modes
corresponds to the (refined) topological vertex.  The oscillator modes of the string theory correspond
to off-shell states and carry information about co-dimension 2 defects of the superconformal theory.
Particular limits of a subset of scattering amplitudes of this string theory lead to
the partition functions of $A_n$ Gaiotto theories for all $n$, compactified on $S^4$, i.e.,
to amplitudes of all Toda theories.

\end{titlepage}
\setcounter{page}{1} 

\section{Introduction}
Protected amplitudes of supersymmetric theories are frequently captured by topological
theories.  These are typically holomorphic quantities.  In the context of supersymmetric theories arising in string theory these are computed by topological string amplitudes.  

With the work of Pestun \cite{pestun} on partition function of  supersymmetric theories on $S^4$, it became clear that not only holomorphic amplitudes of supersymmetric theories
can arise in partition functions but also combinations of topological/anti-topological
amplitudes arise--very much along the lines of $tt^*$ geometry studied in the 2d context \cite{cv}.
This suggests that the notion of a  `real' topological strings exists,
namely combining the usual topological string with its complex conjugate to come up
with a real version.  In many cases we already have all the relevant tools to study the
corresponding amplitudes.  The question is what is the meaning of
the resulting physical system?

In this paper we study in particular the supersymmetric partition functions associated
with 5d superconformal theories which arise by compactification of M-theory
on singular Calabi-Yau manifolds, such as toric 3-folds \cite{dkv,mseib,ims,kmv}. Alternatively these
are described \cite{lv} by the dual type IIB network of $(p,q)$ 5-branes \cite{web}.  The superconformal
index for these theories, i.e. their partition function on $S^1\times S^4$ is
computable using topological string amplitudes on the corresponding Calabi-Yau \cite{iv}.  

Topological string amplitudes on toric CY threefolds can be computed using the topological
vertex formalism \cite{akmv} (and its refinement \cite{ikv}), which involves associating Feynamn like
rules, where the relevant Feynman diagrams correspond to the geometry of the $(p,q)$
5-brane web.  This originally motivated the question of defining a quantum field theory whose amplitudes
will automatically give the Feyman diagrams.  However, there was a notable
difference between Feynman diagrams and the amplitudes for topological strings:
In the case of Feynman diagram one integrates over the internal momenta, one for each
internal loop.  The analog of this for the web diagram would translate to integration
over the moduli space of the corresponding Calabi-Yau, which one usually holds
fixed in the case of topological strings.  

However as we shall see, in computing the superconformal index associated to the resulting
theory, one ends up integrating over the internal loops, one integral
per loop!  So this major difference between the topological string diagrams and
Feynman diagrams disappear and motivates the question:  the Feynman diagrams
are describing perturbation of which quantum theory?  The main aim of this paper is
to propose an answer to this question.  We propose that these Feynman diagrams correspond to
scattering amplitudes of a string theory propagating in 4-dimensional non-commutative
version of space-time  $T^*{\bf T}^2$.  The choice of the 5d superconformal theory translates to the choice
of the winding sectors of the string scattering states and the choice of the transverse
position of the winding string on ${\bf T}^2$ translate to scattering momenta of the Feyman diagrams
which are chemical potentials for global symmetries of the 5d superconformal theory.
Moreover the oscillation modes of the string are off-shell states and they enjoy a natural cubic vertex structure,
which is the (refined) topological vertex.  The off-shell states capture amplitudes
associated with the codimension 2 defect
operators of the 5d superconformal theory.  The natural Wick rotations of this
theory suggests a signature $(2,2)$ for the space-time, and raises the question
of relation of this string theory to $N=2$ strings \cite{ov}.

The corresponding 5d superconformal theories reduced on a circle give a rich
class of ${\cal N}=2$ superconformal theories, which include in particular all the
$A_{n}$ Gaiotto theories \cite{gaiotto}.  Thus the partition functions of these theories,
which are conjectured to be related to Toda theories, get related to specific
scattering amplitudes of the string theory we propose, where the choice
of the scattering states in the string theory, determines the choice of the Gaiotto theory.
It would also be interesting to investigate the properties of the more general class of 4d ${\cal N}=2$ theories that we obtain, whose partition functions are computed by the string theory we propose.

The organization of this paper is as follows:  In section 2 we review the construction
of 5d conformal theories obtained from M-theory on toric manifolds
(including some which admit singular toric action) or equivalently
type IIB on a web of $(p,q)$ 5-branes (which could end on 7-branes).  In section 3 we discuss how one can
use topological strings to compute superconformal indices for these theories.
In section 4 we propose the string theory in 4 dimensions, and in section 5 we end with
some open questions.

\section{Toric Geometry, (p,q) 5-brane Web and 5d Superconformal Theories}

It is believed that compactifying M-theory on singular Calabi-Yau 3-folds
can lead to superconformal theories in 5 dimensions involving interacting
massless particles and tensionless strings \cite{wphase,dkv,mseib,ims}.  It is also known \cite{lv}
that this is dual to a web of $(p,q)$ 5-branes \cite{web}.
In this section we review some basic features of these conformal theories, reformulated
in a way which will be most useful to us in this paper.

A toric Calabi-Yau 3-fold is captured by the geometry of which cycles of a ${\bf T}^2$ shrink
and what their intersection geometry is.  Moreover the duality between M-theory on ${\bf T}^2$
and type IIB on $S^1$ allows us to convert a locus with a $(p,q)$-cycle of ${\bf T}^2$ shrinking
with a $(p,q)$ 5-brane of type IIB.  Thus the geometry of toric Calabi-Yau 3-folds
can be captured by a web of $(p,q)$ 5-branes.  Each such 5-brane fills the 5-dimensional
space-time and has 1-dimension extended along a straight line in ${\bf R}^2$.  The intersecting structure
of the $(p,q)$-branes in ${\bf R}^2$ forms a web $W\subset {\bf R}^2$. We can view
$W$ as oriented, where the orientation on the web distinguishes $(p,q)$ brane
from a $(-p,-q)$ brane.  
Charge conservation
of the $(p,q)$ 5-brane implies that $\sum_i (p_i,q_i)=0$ for each junction (vertex) where
$(p,q)$ 5-branes meet.   The generic
junctions of the $(p,q)$ 5-branes are cubic.  Moreover, if we introduce
the inner product between $w=(p,q)$ 5-branes, given by
$$\langle w_1,w_2\rangle = \langle (p_1,q_1), (p_2,q_2)\rangle =p_1q_2-q_1p_2,$$
the generic cubic vertex has the property that
$$\langle w_1,w_2\rangle =\pm 1$$

\begin{figure}[ptb]
\begin{center}
\includegraphics[
height=4.3215in,
width=4.1361in
]{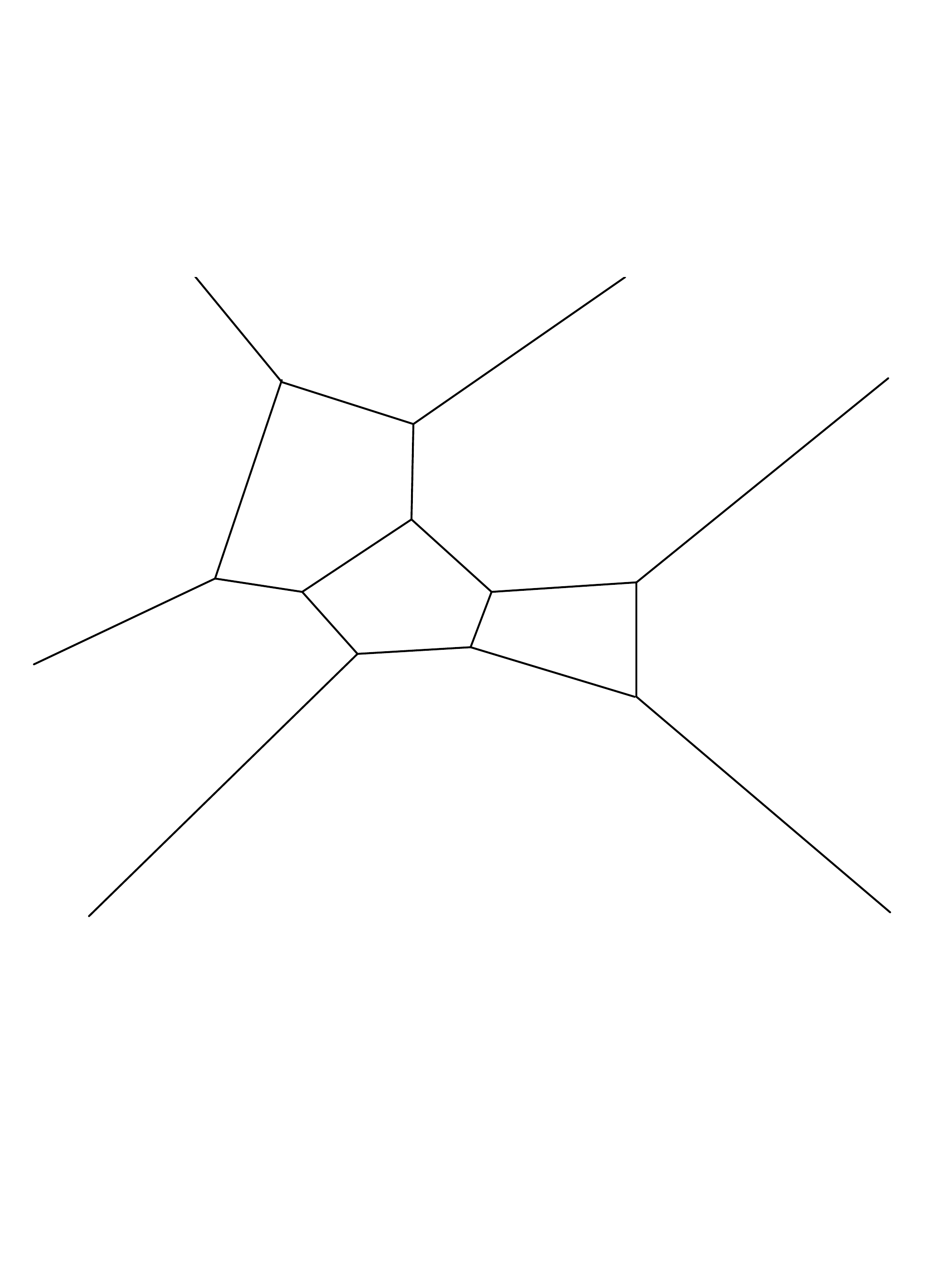}
\end{center}
\caption{{\protect\footnotesize {An example of $(p,q)$ web with $N=6$ external lines and
$g=3$ internal loops.}}}%
\label{breathe}%
\end{figure}

\begin{figure}[ptb]
\begin{center}
\includegraphics[
height=4.3215in,
width=4.1361in
]{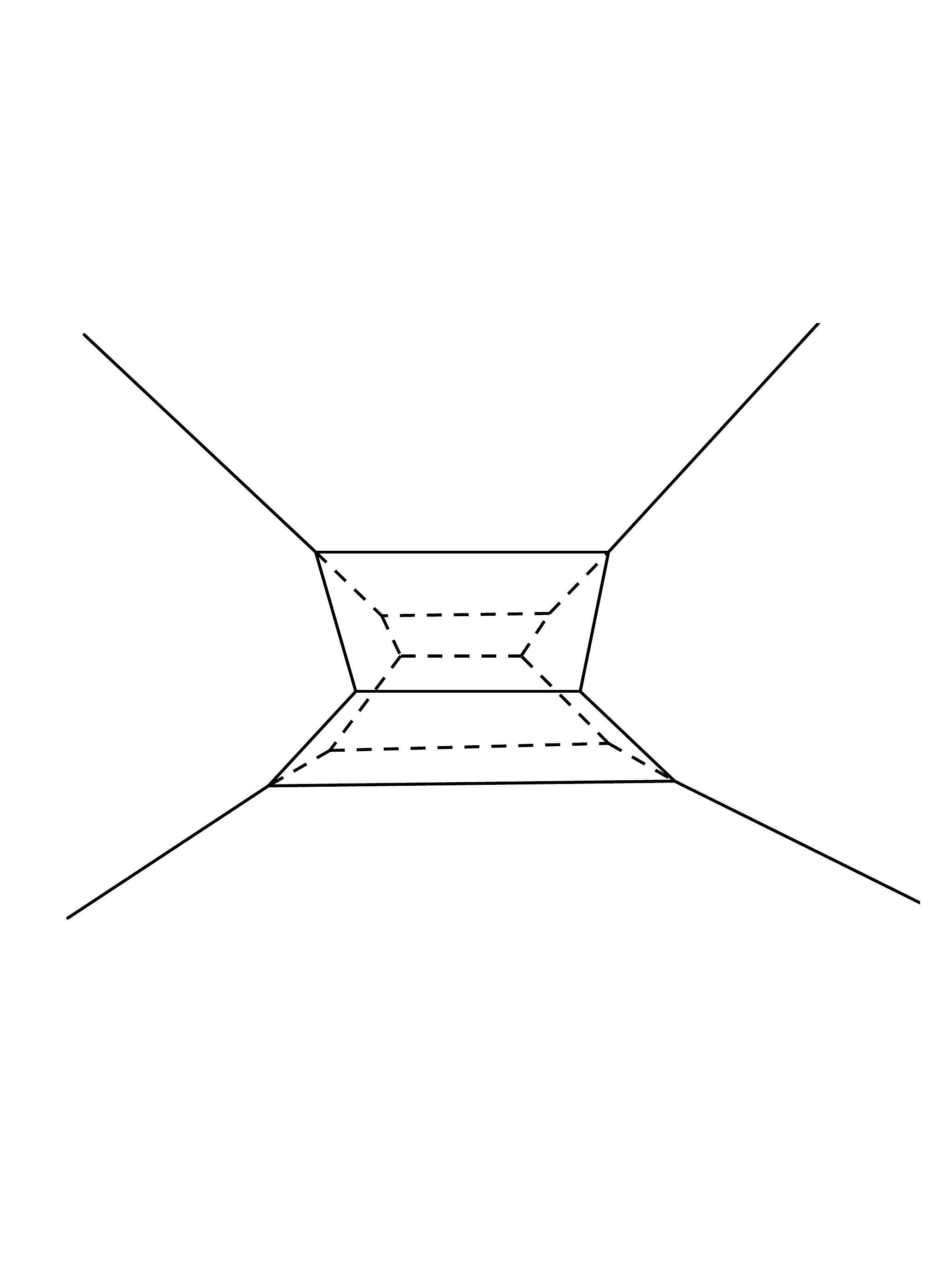}
\end{center}
\caption{{\protect\footnotesize {A breathing mode corresponds
to changing the web, without moving the external lines.  There
is one breathing mode per internal loop.}}}%
\label{etafour}%
\end{figure}
Properties of these theories are nicely captured by the properties of the branes \cite{web}, which
we now review.
   To balance the tension between the branes at the junctions, the orientations
 of $(p,q)$ 5-branes should be such that there is no force at the vertex.  This is automatically
 the case if the $(p,q)$ brane is directed along the line $(p+q \tau )$
where $\tau$ is the type IIB coupling constant.  Up to a rotation in $R^2$ this
is the most general solution for balancing the tensions.  Since $\tau$ will not play
a role in this paper, for simplicity of exposition we take $\tau =i$, in which case
the $(p,q)$ 5-brane is directed along a line with slope $q/p$.
For an example of a $(p,q)$ web see Fig.\ref{breathe}.

Any web $W$ will have a number of $(p,q)$ branes which are semi-infinitely
extended along ${\bf R}^2$.  These we will call the `external lines' of the web.
The rest of the 5-branes will be called `internal'.  
Moving the external lines in ${\bf R}^2$ requires infinite energy compared
to the internal lines.  Thus the degree of freedom of moving the external
lines are non-normalizable and can be viewed as `mass parameters' of the theory.
The internal lines correspond to normalizable modes.  Let us fix the mass parameters, and
thus the position of external lines.  Of course we cannot move
the internal lines arbitrarily.  First, the branes can only move parallel to themselves,
to preserve tension.  Secondly, the external lines are fixed.  It is easy to see that
for each primitive closed cycle (loop) in the web diagram, there is one mode which corresponds
to the `breathing mode' of the loop (see Figure \ref{etafour}).  This is in perfect agreement with the M-theory picture,
as each  closed loop corresponds to a 4-cycles in the toric Calabi-Yau, and the primitive
loops form a basis for 4-cycles.  Moreover
the corresponding breathing mode of the loop is the scalar corresponding to the normalizable K\"ahler mode.
The internal lines of the web $W$ correspond
to 2-cycles of the Calabi-Yau, though they are not independent classes.  We shall denote by $N$ the number of external lines and by $g$ the number of primitive
loops of the web.  The theory will have a $U(1)^g$ gauge symmetry and the breathing
modes are the real scalars in the corresponding vector multiplets.
Note that we can view the web as a Feynman diagram with $N$ external lines,
describing the scattering of $N$ particles at $g$-th loop order.  
In general, for fixed external lines, as we change the K\"ahler classes, the topology
of the web may change, though generically it will still have $g$ primitive loops.
From the viewpoint of Calabi-Yau this corresponds to transitions from
one Calabi-Yau to another.

\subsection{Integral convex polygon associated to the web}

Consider a web of 5-branes with $N$ external lines corresponding to classes $w_i$ where
we orient the external lines towards infinity.
Then by the brane charge conservation we know that
$$\sum_{i=1}^N w_i=0.$$
Using the external lines we can define an integral convex polygon $P$ in the following way:
Consider an infinite radius circle and order $w_i$ in increasing order as we go
around the circle in a counter-clock wise direction.   If there are several
parallel $w_i$, they will clearly be one after the other in this sequence.   

Using this data we can define an integral convex polygon $P$.  This is a polygon
whose vertices are on integral points of ${\bf R}^2$.  Furthermore the edges of $P$ passes through
$N$ integral points.  Let $a_j$ correspond to the $j$-th integral point of $P$.
This polygon is defined (up to an overall shift in the lattice) by the condition that
$$a_{j+1}-a_j=iw_j$$
where $iw_j=i(p_j,q_j)=(-q_j,p_j)$.
Note that since $\sum_i w_i=0$ we learn that $a_{N+1}=a_1$.  In other words we
have a closed polygon passing through $N$ integral points.  The fact that $P$ is convex
follows from the fact that the angle that  $w_i$ makes with the positive $x$-axis is increasing
(in the counter-clockwise sense).   The number of integral points on each edge of $P$ minus one is
the number of parallel 5-branes in that direction.

\begin{figure}[ptb]
\begin{center}
\includegraphics[
height=4.3215in,
width=4.1361in
]{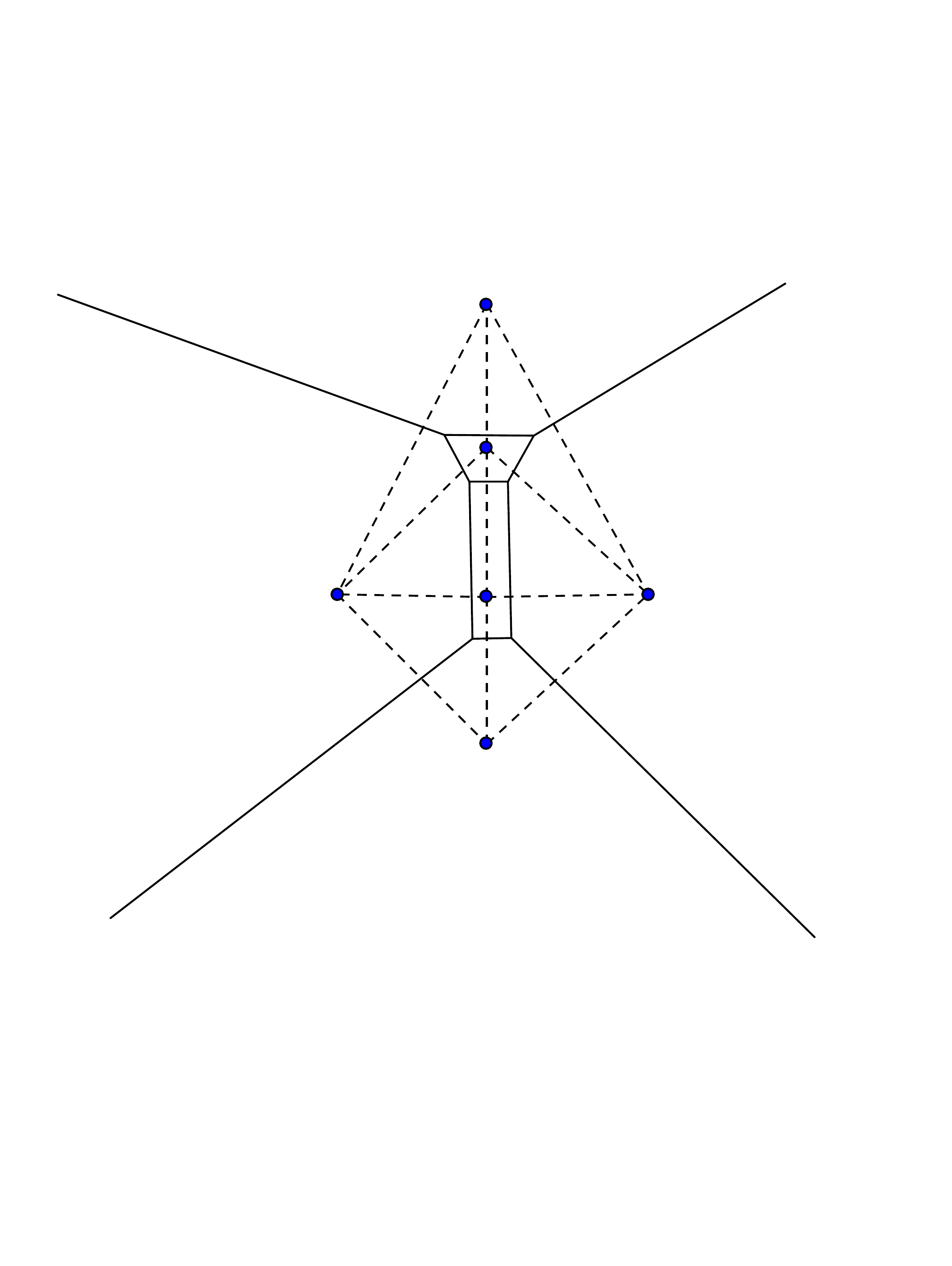}
\end{center}
\caption{{\protect\footnotesize {An example of a $(p,q)$ web with $N=4$ external lines and
$g=2$ internal loops.  The convex outer polygon has 4 sides and its triangulation
is dual to the web.}}}%
\label{etatwo}%
\end{figure}

Even though it is not obvious, $g$ is the number
of integral points in the interior of $P$.  Moreover the number of internal points $g$ can
be computed according to (Pick's theorem):

$$g=A-{N\over 2}+1={1\over 2} \sum_{i<j} \langle w_i,w_j\rangle -{N\over 2} +1$$
where $A$ is the area of $P$.
In fact if we consider a generic web with the asymptotic states fixed by a polygon,
and view the interior points as vertices (leading to a `grid diagram') and triangulate them, each triangulation
corresponds to a choice of web diagram with fixed external lines.  The web is 
dual to this triangulation.
In this picture,
each face of the triangulation corresponds to a cubic vertex of the web and each
edge of the web is orthogonal to an edge of the triangulation.  See Figure \ref{etatwo}. 

If we compactly this 5 dimensional theory on a circle we obtain an ${\cal N}=2$ theory in
4 dimensions.  The $(p,q)$ web of 5-branes can get dualized to an M-theory description
involving M5 branes wrapping an SW curve.  Equivalently, in the geometric
picture of the M-theory on toric geometries we are led to type IIA on
toric manifolds which is mirror to type IIB on the 3-fold with geometry
$F(X,P)=uv$ where the curve $F(X,P)=0$ is the SW curve and is given by
$$\Sigma: \qquad F(X,P)=\sum_{m,n\in P} a_{mn} X^mP^n =0$$
where ${m,n}$ are integral points inside, or on the boundary of $P$.  The $X,P$ are ${\bf C}^*$ variables and we sometimes also write them as
$$X={\rm exp} (x)={\rm exp}(-\tau_x+i\theta_x)$$
$$P={\rm exp}(p)={\rm exp}(-\tau_p+i\theta_p)$$
$X,P$ can be explained in the type IIB parameterization as follows:
The $(\theta_x,\theta_p)$ can be naturally identified with the F-theory elliptic fiber $T^2$ and $(\tau_x,\tau_p)$ denotes the ${\bf R}^2$ on which the web of $(p,q)$ 5-branes were 
stretched.  The $a_{mn}$ are complex parameters which parameterize the Coulomb
branch of the theory if $(m,n)$ is one of the $g$ interior points of $P$.  If $(m,n)$ is on
the boundary of $P$, $a_{mn}$ is a mass parameter of the theory.  The SW differential can be identified
with 
$$\lambda_{SW}=pdx.$$
The SW curve $\Sigma$ can be identified as the fattened up version of the web diagram.
Moreover if $(m_1,n_1)$ and $(m_2,n_2)$ correspond to two successive points
on the boundary of $P$ the cylinder associated with that edge comes from the
part of the polynomial given by
$$a_{m_1n_1}X^{m_1}P^{n_1}-a_{m_2n_2}X^{m_2}P^{n_2} \sim 0$$
leading to the cylinder given by
$$X^{m_2-m_1}P^{n_2-n_1}=a_{m_1n_1}a^{-1}_{m_2n_2}$$
which corresponds to the $(p,q)$ 5-brane associated with $w=(n_2-n_1,m_1-m_2)$.

\subsection{Adding 7-branes}

\begin{figure}[ptb]
\begin{center}
\includegraphics[
height=4.3215in,
width=4.1361in
]{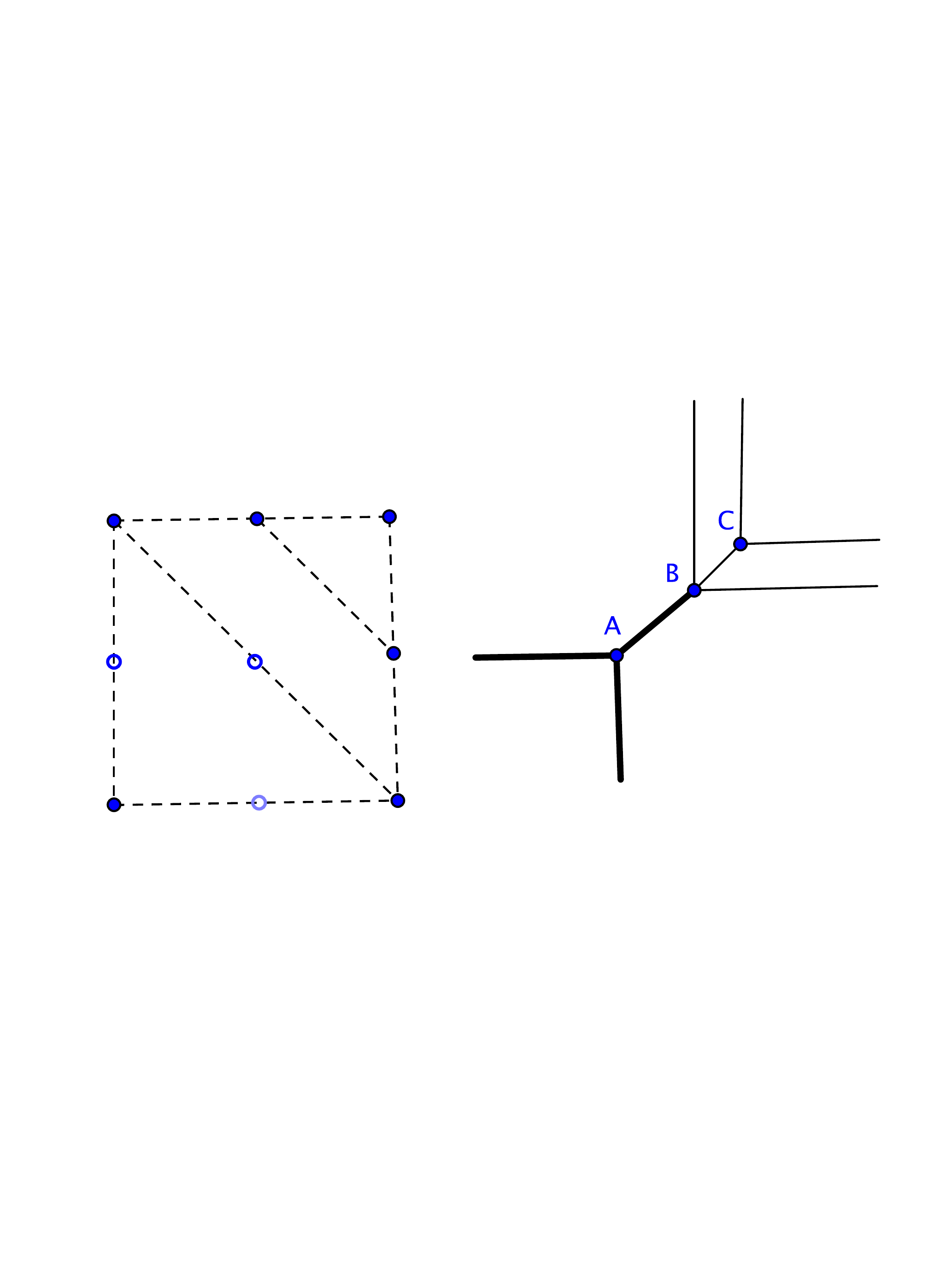}
\end{center}
\caption{{\protect\footnotesize {An example of grouping 5-branes according to which
7-branes they end on.  The dual grid contains white dots to depict the corresponding
grouping.  The vertices of the web correspond to faces of the grid.  Vertex A corresponds
to the bigger triangle.  Vertex B the trapezoid and vertex C to the small triangle.  The
thicker edges of the web correspond to two 5-branes.}}}%
\label{etathree}%
\end{figure}

It is also possible to add 7-branes to this story, where the 5-branes can end on.
This has been recently revisited in \cite{tach}.
These will correspond to M-theory compactified on non-toric Calabi-Yau which admit singular torus actions.  
This construction will allow us to also discuss manifolds such as ${\bf P}^2$ blown up at
8 points inside a Calabi-Yau in the context of webs.  In this context it is natural to put the 7-branes
at infinity in ${\bf R}^2$.  Thus the external lines will be grouped according
to which 7-brane it ends on.  Sometimes this geometry is depicted by adding
white dots to the diagram \cite{tach}.    The white dots signify the grouping
of the 5-branes.  In particular the 5-branes ending on the same 7-brane
are dual to edges which are separated by white dots.  See Figure \ref{etathree}.
The rules of what type of vertices the 5-branes with multiplicity can have,
has been worked out and we review them later in section 4.

\subsection{Introducting defects}

\begin{figure}[ptb]
\begin{center}
\includegraphics[
height=4.3215in,
width=4.1361in
]{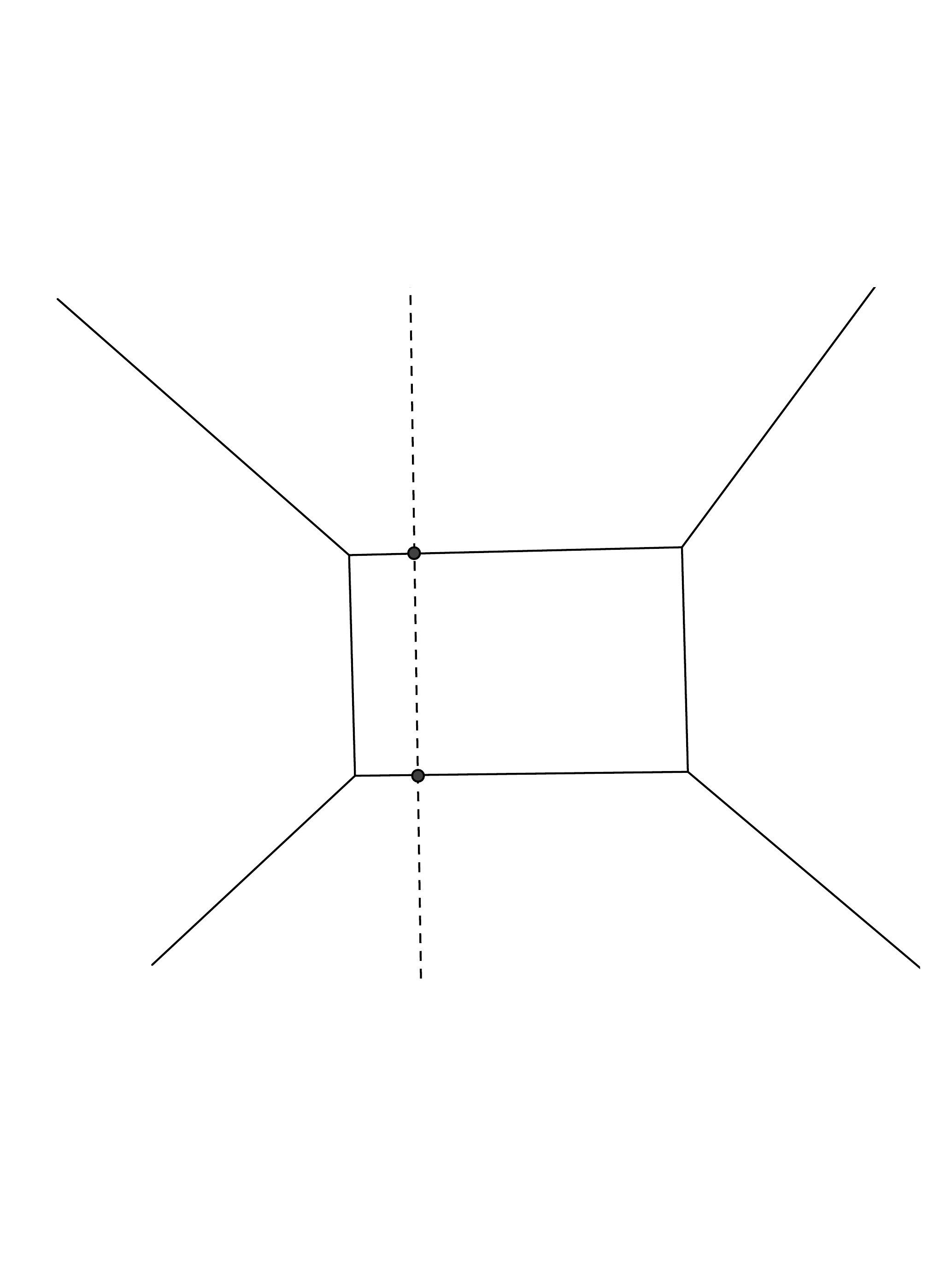}
\end{center}
\caption{{\protect\footnotesize {The D3 branes (or Lagangian branes in
the M theory setup), are suspended between a 5-brane at infinity
(denoted by a dashed line) and the web.  The D3 branes are perpendicular
to the plane ending on the web at one of the two dots.}}}%
\label{defect}%
\end{figure}

It is also possible (and important) to introduce defect probes in this theory.  In the M-theory
setup on Calabi-Yau, they correspond to wrapping a number of M5 branes on 
a special Lagrangian subspace  \cite{MV}, filling a 3-dimensional subspace of space-time.  This can be viewed as a codimension 2 defect in the 5-dimensional theory.
In the language of $(p,q)$ 5-brane web, they correspond \cite{akv} to introducing D3 branes
which fill a 3-dimensional part of 5d space-time and correspond to a semi-infinite
line in the orthogonal direction to the ${\bf R}^2$ plane of the web, which
end on an edge on the 5-brane.  To fix a class of such defect probes it turns out
to be natural to view the semi-infinite D3 brane probe to be a finite segment that ends on a spectator $(p,q)$ 5-brane
which is placed infinitely far above the plane of the web (see Figure \ref{defect}).  In the geometric
language introducing the spectator 5-brane at infinity corresponds to choosing a compactification
of the special Lagrangian which the M5 brane wraps.

If one compactifies this theory on a circle down to 4 dimensions, the resulting defect
will correspond to a surface operator of the ${\cal N}=2$ theory.  In the context of
webs which give rise to gauge theories some of these probes can be viewed as surface operators
of the type studied in \cite{gw}  and connected
to this picture in \cite{wk,dg}.  They correspond to choosing a collection
of points on the SW curve, where the D3 brane intersects the web.

In this context as we consider the normalizable deformations of the web, i.e. the breathing
modes associated with the closed loops in the web, the spectator 5-branes at infinity
do not move, but the D3 brane suspended between the spectator 5-brane and
the web slide along the spectator brane  in unison with the breathing mode, so as to continue ending on the web.  This defines the class of the defect operators as we give expectation
values to the fields.  It is in this context that it becomes clear that we need the spectator
$(p,q)$ 5-brane at infinity to actually define the defect operator.  Different choices
of the spectator 5-brane lead to different defect operators.  Needless to say we can introduce
as many defect probes and as many D3 branes suspended between them and the web,
as we wish.  The geometry of where the D3 brane probe ends on the web, defines different
types of defect operators, which map, upon reduction to 4d in the gauge theory case,
to the different types of defects allowed in gauge theory.

\subsection{5d Superconformal Points}

\begin{figure}[ptb]
\begin{center}
\includegraphics[
height=4.3215in,
width=4.1361in
]{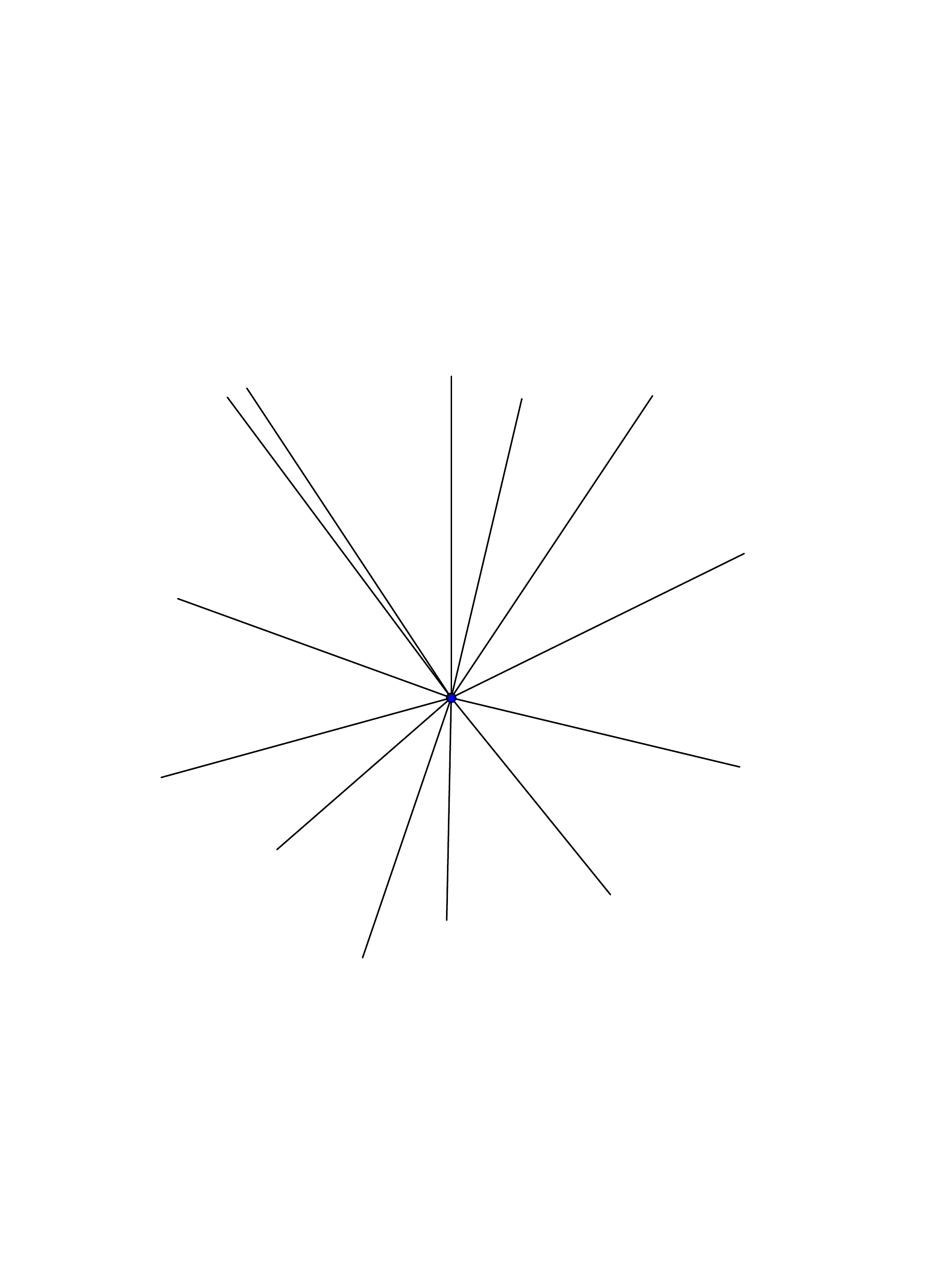}
\end{center}
\caption{{\protect\footnotesize {At the conformal point the web
collapses to rays passing through one point.}}}%
\label{point}%
\end{figure}
When 4-cycles in the CY shrink to zero size, we get tensionless strings
by wrapping M5-branes over them.   In the type IIB setup the
tensionless string corresponds to D3 branes wrapping
the minimal cycles and ending on the (p,q) 5-branes. In addition M2 branes wrapping over
the corresponding 2-cycles inside the 4-cycle is massless.  We thus get
a collection of interacting massless particles and tensionless strings which suggests
we have a conformal fixed point.  For each web, with at least 1 internal
integral point (i.e. $g\geq 1$) we thus end up with a non-trivial conformal theory
where all the internal loops vanish.   In this limit the web collapses to $N$ semi-infnite
lines passing through the same point (where some semi-infinite lines may
have multiplicity, associated to external 5-branes with the same $(p,q)$). 
See Figure \ref{point}. We
can further group these according to which parallel ones end on the 7-brane.
It is natural to conjecture that for each such configuration there is a unique superconformal
theory in 5 dimensions.  Thus we can associate to each integral convex polygon
$P$, with possible white dots which signify the grouping of the same type $(p,q)$ branes,
a superconformal theory.   Two convex integral
polygons define the same superconformal theory if they
can be mapped to one another by translation on the integral lattice, as well as an
$SL(2,Z)$ change of basis.

 Note that the number of mass parameters for this
conformal theory is $N-3$ where $N$ is the number of (black) integral points on the boundary
of the polygon (i.e. external lines).  This is because we have one mass parameter
corresponding to moving each group of external lines, except that a two parameter subspace
of this simply corresponds to moving the intersection point of the lines on ${\bf R}^2$ and does not
change the theory.  We first need to say more precisely how the masses are parameterized:
For each brane given by a vector $w\in H_1(T^2,{\bf Z})$ we choose a $v \in H_1(T^2,{\bf Z})$
such that $\langle w,v\rangle =1$.  $v$ is unique up to shifting it by an integer multiple of $v$:
$$v\rightarrow v+nw$$
Each such choice of $v$ we will call a `framing' associated to the brane $w$.  Consider
moving the brane $w$ along ${\bf R}^2$.  We parameterize the motion by a real number (mass)
$m\in {\bf R}$ multiplied by $v$.  In other words, we move every point of $w$ on ${\bf R}^2$
by $mv$.  Note that the choice of framing is irrelevant for this motion as different
framings correspond to shifting the brane parallel to itself which does not change the brane.

We now show that if we parameterize the $m_i$ corresponding to movement of the
branes from the superconformal point (i.e. where $m_i=0$ correspond to the superconformal
point) then they are not independent.  In particular
$$\sum_{i=1}^N m_i =0.$$
To show this 
we will first show this for three primitive brane junctions.
Consider the case with three primitive 7-branes, $w_1,w_2,w_3$
with
$$\sum w_i=0 \quad and \quad \langle w_1,w_2\rangle =\langle w_2,w_3 \rangle =\langle w_3,w_1\rangle =1$$
Note that by an $SL(2,Z)$ transformation we can take $(w_1,w_2,w_3)=((1,0),(0,1),(-1,-1))$.
We can conveniently choose the framings
$$v_1=w_2, v_2=-w_1,v_3=w_1$$
If we move $w_1$ by $m_1v_1=(0,m_1)$ and $w_2$ by $m_2v_2=(-m_2,0) $ the intersection point between
$w_1$ and $w_2$ branes 
will have moved by $(-m_2,m_1)$.  Now the $w_3$ brane will have to be
moved to pass through it.  Indeed if we shift $w_3$ by $m_3v_3=(-m_1-m_2)(1,0)$ then
it passes through it because $(-m_2,m_1)-(-m_1-m_2,0)=m_1(1,1)$.  We thus learn that
consistency of the brane configuration requires that
$$m_1+m_2+m_3=0$$
Now let us consider the general web in a generic position\footnote{Here we assume there is only a single 5-brane on each edge; this
argument can easily be extended to the case where we have groups
of $(p,q)$ branes ending on 5-branes.  In this case, as we will
mention in section 4 two types of vertices enter,
a cubic and a quartic, for each of which the same argument gives the conservation law.} .  We can associate to each
edge a mass parameter $m_i$.  The generic web will have cubic vertices
which as we have just shown require $\sum m_i=0$ at each vertex.  Thus
the conservation of $m_i$ implies that the sum of the $m_i$ over all the external
lines vanishes:
$$\sum_{i=1}^N m_i=0,$$
which is what we wished to show.   Moreover, we can show that two combination of mass
parameters are redundant.  In fact let $u_1,u_2$ be a basis for
$H_1(T^2,{\bf Z})$.   Translations of ${\bf R}^2$ by $u=r_1u_1+r_2u_2$ will shift all the masses according to
$$m_i \sim m_i+\langle w_i,u\rangle $$
Thus out of $N$ potential mass parameters we are left with $N-3$ independent parameters.

Upon compactification on a circle the mass parameters become complexified
and as we have discussed they are captured by the ratio of the coefficients associated
with the boundary monomials of $P$.  In particular to each external line $w$ we associate the
exponentiated
mass parameter
$$M_w= exp(-m_w+i \theta_w)$$
Furthermore, recall that each $w$ is 
associated with a boundary edge $w=-i(m,n)=(n_2,-m_2)- (n_1,-m_1)$ of $P$ and we get the
identification
$$M_w =a_{m_1n_1}/a_{m_2n_2}$$
The analog of the mass condition $\sum m_i=0$ gets mapped to
$$\prod_{w\in \partial P} M_w =1$$
which is automatic.  Moreover rescaling $X\rightarrow \alpha X$ and $P\rightarrow \beta P$ shifts
$$M_w\rightarrow M_w\ \alpha^m \beta^{n}$$
which gets rid of two additional complex mass parameters.  In other words the angular
part of the mass parameters $\theta_w$ also satisfy the constraint
$$\sum \theta_w =0$$
and
$$\theta_w \sim \theta_w+ \langle w, u\rangle$$
These can be interpreted as follows.   Let us consider
one of the external lines given by $w$.  For simplicity of exposition let us take
the external line corresponding to $w=(1,0)$.  This corresponds to tube spanned
by $X$.  In general we can parameterize the SW curve by choosing
$X$ as the local coordinate.  In this way $P$ can be viewed as a holomorphic
function $P(X)$ along the tube.  More precisely we can
view $p(x)dx$ as a 1-form  along the tube which gives the data of the geometry.
 For $|X|>>1$ the transverse position of the tube approaches a constant on the $P$ tube given by
$$P=M_w$$
\begin{figure}[ptb]
\begin{center}
\includegraphics[
height=4.3215in,
width=4.1361in
]{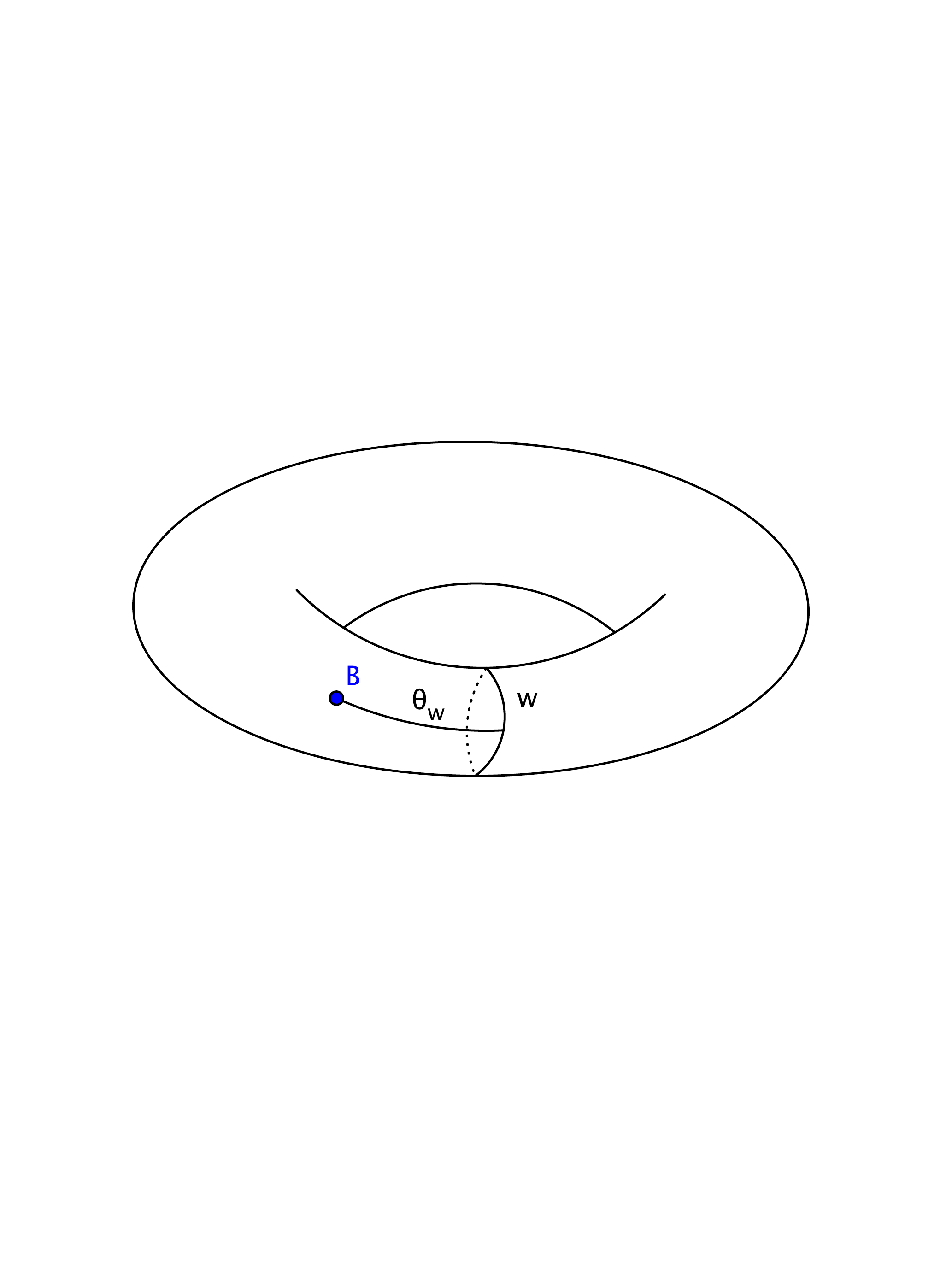}
\end{center}
\caption{{\protect\footnotesize {An external line is characterizes by a winding $w$
on ${\bf T}^2$.  The transverse position of the string, as measured from some
base point $B$ on ${\bf T}^2$ corresponds to $\theta_w$.}}}%
\label{circle}%
\end{figure}

Changing $m_w$ corresponds to moving the position of the transverse
point of the tube along $\tau_p$ and changing $\theta_w$ corresponds to rotating
the position of the tube in the angular direction.   Note that if we project this tube
to the ${\bf T}^2:(\theta_x,\theta_p)$, it is wrapped around $\theta_x$ at a position given by $\theta_p=\theta_w$.
The angular part of the framing of a tube depends
on which base point we measure the position of $\theta_x$ from.  Shifting the
base point on ${\bf T}^2$, where we measure $\theta_w$ corresponds to the degree of freedom of shifting the angular
part of the masses by two parameters.  See Figure \ref{circle}.

\section{Topological Strings and 5d Superconformal Indices}
Consider M-theory on a Calabi-Yau threefold $K$.  In addition compactify
further on $S^1\times TN$ where $TN$ denotes Taub-NUT.  The circle
product is twisted: as we go around $S^1$ we rotate the two
complex planes of $TN$ by 
$$(z_1,z_2)\rightarrow (qz_1,q^{-1} z_2)$$
If the Calabi-Yau is non-compact as in toric cases, and has an extra geometric
$U(1)$ rotation corresponding to R-symmetry, we get a refinement of this
by allowing independent rotations
$$(z_1,z_2)\rightarrow (q_1z_1,q_2^{-1}z_2)$$
and rotating the internal geometry by a $U(1)$ rotation given by $q_1q_2^{-1}$.
The partition function of M-theory for this geometry is a definition of topological
string \cite{dvv}:
$$Z_M(K\times S^1\times_{(q_1,q_2)} TN)=Z^{top}(K,q_1,q_2)$$
A series of arguments \cite{dvv} relate this definition to the usual definition of A-model
topological string when $q_1=q_2=q=exp(i\lambda)$, where $\lambda$
is the topological string coupling constant.    In this paper for simplicity
we sometimes specialize discussion to this case, though the generalization of the amplitudes
and methods for computation of refined topological strings are also known.

The topological string partition function for toric geometries (which correspond
to webs with multiplicity 1 on each edge) can be computed using topological
vertex formalism \cite{akmv} (and its refinement \cite{ikv}--see also \cite{awata}) which was inspired by the
large $N$ dualities of topological strings \cite{GV}.
  
  As discussed already, the Calabi-Yau on toric geometries
  can be represented by a web $W$ of $(p,q)$ 5-branes.  The generic
such web has cubic vertices.  The topological vertex formalism associates
a Young diagram, or equivalently a finite representation $R$ of $U(\infty)$ to each internal line.  Furthermore to each edge one
associates a `propagator factor' given by $Q^{n(R)}={\rm exp} (-n(R)t)$ where $n(R)$ is the number of
boxes of $R$, $Q=e^{-t}$ and $t$ is the complexified length of the edge (i.e. the integral
$t=\oint_{tube} pdx$).  Note that to obtain
finite amplitudes we need to put trivial representation on the external lines, since $Q_{external}=0$
due to infinite length of external lines.
Moreover for each cubic junction one introduces a vertex
$C(R_1,R_2,R_3; q_1,q_2)$\footnote{There are subtleties we are suppressing
here, having to do with the choice of framing as well as the choice of preferred
direction in the refined case.   See the original literature \cite{akmv,ikv} for more detail.  See
also the recent work \cite{onek,ags}.}.  The partition function is computed by summing over all
representations $R$ for the internal edges.

The fact that each edge is associated with a representation $R$ is reminiscent of the
oscillator mode of a single chiral boson or a complex chiral fermion.  This analogy
proved important in explanation of the structure of the topological vertex \cite{akmv,IH}, as we
will now review.
Consider a tubular geometry of the web.  Let us assume, with no loss of generality,
that the tube corresponds to wrapping the ${\bf C}^*$ direction corresponding to $X$.
The embedding of this tube in ${\bf C}^*\times {\bf C}^*$ can be characterized
by how $P$, the position on the orthogonal ${\bf C}^*$, varies over the tube parameterized by $X$.
Noting that the normal direction $p$  belongs to the cotangent of $x$-space, it is natural
to describe the normal deformation as being a 1-form in the $x$-space.  We thus consider
the normal deformation as
$$pdx=p_{cl}(x)dx$$
where $p_{cl}(x)dx$ can be read from the classical geometry of the mirror curve $F(X,P)=0$,
by solving $p$ in terms of $x$.  It is natural to ask about fluctuations of this classical geometry,
in a holomorphic way.
This in particular is parametrized as
$$p=p_{cl}(x)+\sum_n \alpha_n X^{n}=p_{cl}(x)+\sum_n \alpha_n {\rm exp}(-n(\tau_x+i\theta_x))$$
given that $pdx$ is a 1-form, and it is chiral, it is natural to identify it with a chiral bosonic current:
$$pdx=\partial \phi$$
which satisfies ${\overline \partial}\partial \phi=0$.  In fact more is true:  The radius
of the boson (for the unrefined topological string) is such that the boson is equivalent to a complex fermion and we can write\footnote{It is natural to expect that for the unrefined case
the boson acquires a background charge, though this has not been worked out.}
$$\partial \phi=\psi^* \psi$$
For this to be true, the radius of $\phi$ should be 1, such that $\psi={\rm exp}(i \phi)$. 
This turns out to be true as explained in \cite{IH}.  The basic idea is
that the $\psi$ corresponds to creation operator for the Lagrangian brane in the Calabi-Yau,
or in the type IIB setup as the D3 brane ending on the web.  In other words if we add a codimension
2 defect to our theory, this is equivalent to insertion of $\psi$ at a particular point on the curve.
The period of $\phi$ is precisely correct to capture the charge of the brane ending on the web.

\begin{figure}[ptb]
\begin{center}
\includegraphics[
height=4.3215in,
width=4.1361in
]{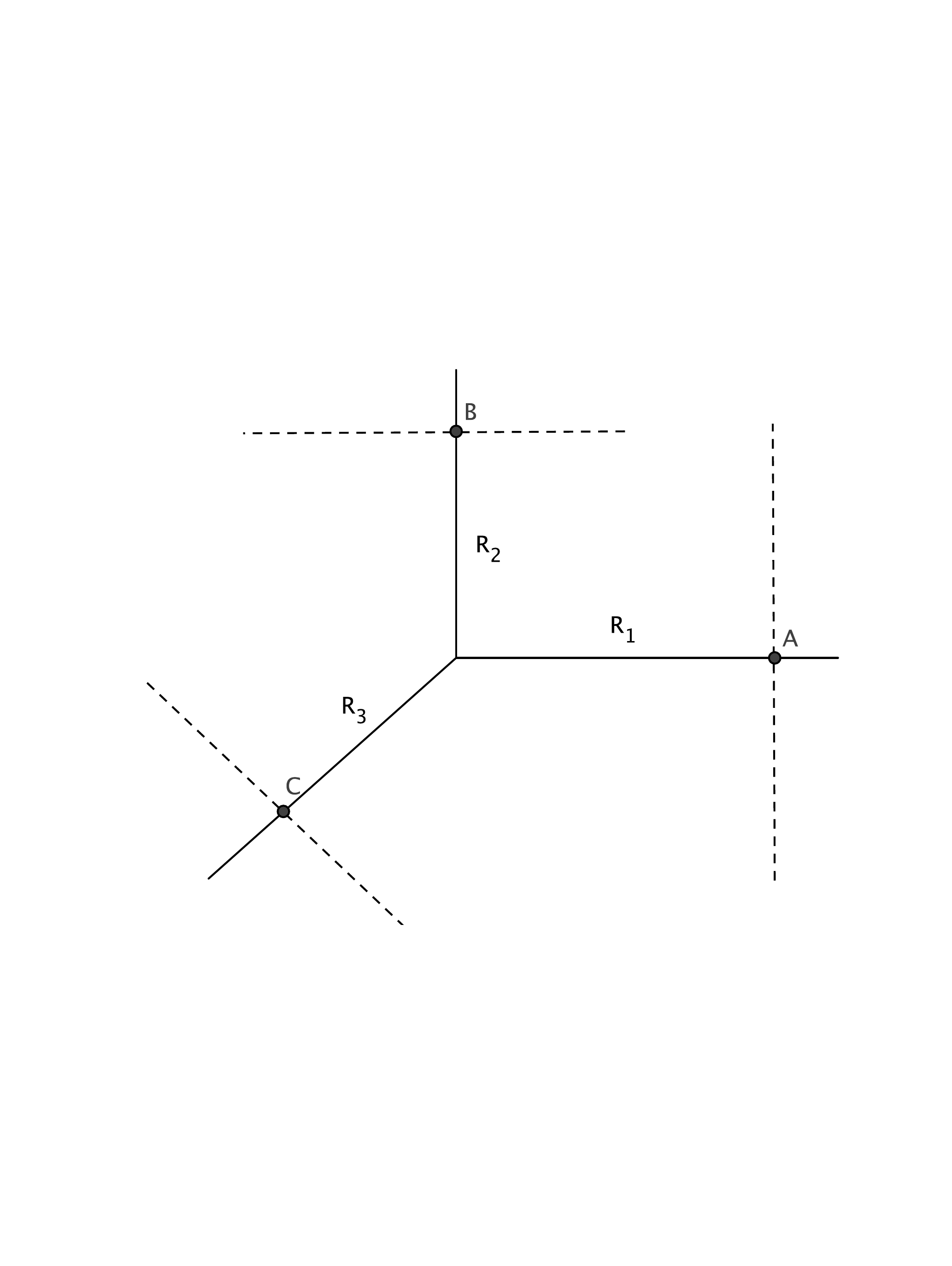}
\end{center}
\caption{{\protect\footnotesize {
Topological vertex can be viewed
as the partition function of topological string in the presence of defects
introduced by branes which end on the web at $(A,B,C)$.  The excitations $(R_1,R_2,R_3)$ on the legs
are created by them.
}}}%
\label{topol}%
\end{figure}

The introduction of defects plays a key role in the definition
of topological vertex \cite{akmv}.  In fact the topological vertex can be viewed as the
study of the configuration dual to ${\bf C}^3$ where we have introduced
three sets of defect probes, one for each edge of the associated web.  In this
way the branes suspended between the spectator 5-brane and the web, create
excitations on each of the corresponding legs of the web, whose amplitude is given by
$C_{R_1R_2R_3}$.  See Figure \ref{topol}.

This structure naturally explains why each edge is labeled by a Young diagram, which
is naturally identified with bosonic oscillator fock space (or equivalently $U(1)$ neutral
states in a  complex fermionic fock space) where the propagator is $Q^{L_0}$.  However
if this description is correct then one would expect that the topological vertex is simply
the vertex associated with a chiral boson on thrice puncture sphere.  This is almost true!
However the topological vertex depends on the couplings $q_i$.  The way
this appears is that $\psi(x)$, the brane creation operator, is not an ordinary fermion on a Riemann surface
$F(X,P)=0$, but rather $x,p$ do not commute when acting on $\psi$.  In particular
in the case $q_1=q_2^{-1}={\rm exp}(i\lambda)$
$$[x,p]=i\lambda$$
We view $F(X,P)=0$ (with suitable ordering of $x,p$) as an operator which
annihilates $\psi$ 
$$F(\hat X,\hat P)\psi =0$$
and this leads to Ward identities which fix the cubic vertex \cite{IH}.
The appearance of this structure was further explained in \cite{sulk} using
the M-theory definition of topological strings, and by applying various
dualities to it.  In particular it was shown there that the topological strings partition function
gets mapped to the partition function of type IIA strings for the flat background on ${\bf R^3}\times {\bf C}^* \times {\bf C}^* \times {\bf R'}^3$, where a D6 brane wraps
${\bf C}^*\times {\bf C}^* \times {\bf R'}^3$ and a D4 brane wraps ${\bf R}^3\times \Sigma$,
where $\Sigma \subset {\bf C}^*\times {\bf C}^*$ is given by $F(X,P)=0$.
Notice that the D4 and D6 brane intersect over $\Sigma$ and the strings stretched
between them leads to a complex chiral fermion on $\Sigma$.
Moreover an NS B-field is turned on over the D6 brane given by
$B=dx\wedge dp/i\lambda$.  Only the holomorphic part of this was relevant
for the topological string, but of course reality of $B$ demands that
$$B={1\over i\lambda} [dx\wedge dp-d{\overline x}\wedge d{\overline p}]$$
This B-field makes the D6 brane non-commutative \cite{sw} with non-commutativity
$$[x,p]=i\lambda$$
Moreover the D4-D6 strings form a representation of this non-commutative
algebra \cite{kw}, making the fermions behave as a D-module, exactly as was originally found for
the topological vertex.  

\subsection{A purely 4 dimensional perspective}

From the description of the topological string amplitudes in terms of D4 and D6 branes
discussed in the previous section, as well as the description given in the case of the
5-brane web, it is clear that the main object of interest takes place in the 4-dimensional space
$$M={\bf C}^*\times {\bf C}^*$$
Getting rid of the ${\bf R^3}\times {\bf R'}^3$ part of the type IIA string,
the D6 brane is replaced by a D3-brane wrapping all of $M$, and D4 brane is
replace by a D1-brane wrapping a holomorphic curve $\Sigma$ in $M$.  Moreover there
is a $B$-field turned on in $M$ which is
$$B={1\over \lambda} {\rm Im} \Omega$$
where $\Omega$ is the holomorphic 2-form on $M$ given by
$$\Omega ={dX\wedge dP\over XP}= dx\wedge dp$$
Thus we have one 4-dimensional and one 2-dimensional brane
in $M$.  We are thus naturally led to ask if there is a theory
on $M$ which is equivalent to topological string associated to toric Calabi-Yau.
The answer turns out to be yes!  Indeed we can consider topological strings directly on $M$.
Since $M$ is hyperK\"ahler the branes can be distinguished in more refined ways,
depending on which complex structure one considers.  The D3 brane, together with the
$B$ field, turns out to be a co-isotropic brane of the type introduced in \cite{ko}
and studied in the hyperK\"ahler context in \cite{kw}.  For a co-isotropic
brane we need $k\wedge k=B \wedge B$ and the $k\wedge B=0$, both of which are
easily checked to be true, where $k$ is the K\"ahler form which we take to be
$$k={i\over \lambda } (dx\wedge d{\overline x} +dp\wedge d{\overline p}).$$
In fact this is a `canonical co-isotropic brane', or `cc brane' (since the $B$ field is identified with the imaginary
part of the canonical holomorphic 2-form on $M$).  This is an $(A,B,A)$ brane
distinguishing its brane type relative to the three different complex structures.
The brane associated with $\Sigma$ is a holomorphic brane relative to the usual
complex structure, but corresponds to a Lagrangian brane relative to the other two complex
structures (as real and imaginary parts of $\Omega$ vanish on it), and it is thus called
a $(B,A,A)$ brane.  In this setup, one case ask what is the nature of the string stretched
between the 4-brane and the 2-brane and that turns out to be a chiral fermion \cite{kw}.
Moreover the canonical co-isotropic brane has a non-commutativity induced by $B$
and fermions become what is called a D-module for this non-commutative algebra
exactly as was expected for the topological strings.  We can summarize this by
the statement that {\it the partition function of the topological string on toric Calabi-Yau 3-fold
is the same as the partition function of topological string on the 4-manifold} $M$ {\it with
one cc brane filling $M$ and one} $(B,A,A)$ {\it 1-brane wrapping the curve} $\Sigma$ {\it mirror to the toric 3-fold, where the coupling constant of string theory is inversely
proportional to the strength of the B-field
on} $M$.

\subsection{Superconformal Index}

The index of ${\cal N}=1$ superconformal theories in 5d, can be computed by studying the partition function
of the theory on $S^1\times S^4$ where we twist $S^4$ by rotations $q_1,q_2$ as we go
around $S^1$.  Moreover we do an R-symmetry rotation by $(q_1q_2)^{-1}$  For some examples of computation of superconformal index see \cite{kimyeong}.
This can be connected with topological strings setup where roughly speaking we can view $S^4$ as gluing two copies of $TN$
and the circle actions on the two $TN$'s are opposite.  This can be viewed
as the partition function of the resulting $4d$ theory on $S^4$.  In this context, as studied
in \cite{pestun} we know that the partition function is given by the integral
of the form 
$$Z_{S^4}=\int  \prod_i d{a_i} \big| Z^{Nek.}(a_i; M_j,q_1,q_2)\big|^2 $$
where the integral is over a Lagrangian subspace of the electric Coulomb parameters $a_i$,
and $M_j$ denote the mass parameters and coupling constants.
On the other hand the Nekrasov partition function for theories coming from 5d, is believed to be equal to the refined
topological string partition function.  We thus learn that
$$Z_{S^4\times S^1}=\int \prod_i da_i \big| Z^{top} (a_i;\theta_j,q_1,q_2) \big|^2$$
This structure is explained in \cite{iv}.  As discussed there, the integration cycle of moduli
of the Coulomb branch is identified with the $a_i=\theta_i$, i.e., the compact imaginary
part of Coulomb branch parameters and $\theta_j$ above denotes the imaginary
part of mass parameter, which is chemical potential associated
with the corresponding global $U(1)$ symmetry. We have one such parameter for each closed loop
in the web diagram.
What this means in the topological vertex
formalism is that, we will have two Young diagrams on each leg, one for the topological
theory and one for the anti-topological theory.  This can be interpreted as making
the chiral boson (or equivalently the chiral fermion) living on $\Sigma$  become a non-chiral 
boson (or fermion).  Moreover the propagators are now
$$Q^{L_0}\ {\overline Q}^{\overline L_0}$$
Furthermore for each edge which is part of the loop we integrate over the angular part
of the breathing mode.  In particular for each loop with integral variable $\theta$ we get a factor
$$\int ... \ \prod_{i\in \ edge\ in \ loop} {\rm exp}(i\theta (L^i_0-{\overline L^i_0}))\ d\theta ...$$
This is similar to what we encounter in string diagrams, with two major differences:
The propagators are associated to real integrals, which corresponds to 
twisting the tube rather than lengthening it (the real part of external mass parameters set
the length of the propagator tubes).  Secondly the angular integral does not
project to level matching condition for left and right movers for each edge, because
the integral includes all edges in a given loop.  This integral projects to 
the condition that the sum of left-moving oscillator numbers for all edges in a given loop
is equal to the corresponding sum over the right-moving oscillator numbers.

As already discussed the superconformal theories we have considered are parameterized
by the following data:  We choose $N$ elements  each of which is an element $w_i\in H_1({\bf T}^2,{\bf Z})$.  More precisely
we choose a specific straight winding mode in each given homology class, which comes
in addition with an angle corresponding to moving the winding mode parallel to itself, which
we denote by $\theta_i$.  We denote the corresponding superconformal index for this
theory by
$$I(w_1,\theta_1,...,w_N,\theta_N; q_1,q_2)$$
As we have already noted $I$ vanishes unless the following
conservation laws hold:
$$\sum_{i=1}^N w_i=0, \qquad \sum_{i=1}^N \theta_i=0$$
In other words
$$I(w_1,\theta_1,...,w_N,\theta_N; q_1,q_2)=\delta(\sum w_i)\delta (\sum \theta_i) f(w_i,\theta_i;q_1,q_2)$$
We have seen evidence of a new string theory in 4-dimensions.  We now collect
these evidences and propose a new string theory in 4 dimensions.  

\section{A String Theory in 4 Dimensions}

The string theory we propose propagates in the 4-dimensional space
$$M={\bf C}^*\times {\bf C}^*=T^* {\bf T}^2.$$
This space comes equipped with a natural flat K\"ahler metric.  This
would be viewed as the Euclidean version of the theory.  There is no
natural $(3,1)$ signature version of this theory, due to the importance
of symplectic/holomorphic structure in this geometry.  The closest analog
of Wick rotation in this theory, due to its complex structure, would correspond to changing
the signature of the metric in the ${\bf R}^2$ in the cotangent directions of $M$, $(\tau_x,\tau_p)$.  
In this way each of the asymptotic tubes will in fact be stretched in the usual $1+1$ signature
geometry.  Thus the Wick rotation would be associated to a $(2,2)$ signature space, as is the case with
$N=2$ strings \cite{ov}\footnote{In the $N=2$ string case both of the transverse oscillators
to the worldsheet ends up being a gauge degree of freedom.  This is related
to the fact that the $N=2$ worldsheet has in addition a $U(1)$ gauge field on the worldsheet.}.

We will now describe its physical on-shell states, its off-shell states,
and its scattering amplitudes.  We will then suggest some steps in how one can
come up with a more direct definition of such a string theory.
A prominent role will be played by the ${\bf T}^2\subset M$, which we think of
as spatial directions for this string theory.
The Hilbert space of the theory involves winding strings
which we parametrize by a winding direction $w$.  The on-shell
string states will be identified with straight winding strings which in addition
are labeled by an angle $\theta$ denoting the transverse position of the winding
string as measured from a fixed point on ${\bf T}^2$.  The oscillator modes of the string, which we view
as `off-shell' states, correspond to transverse oscillation modes, one left-moving
and one right-moving which can be classified by a pair of Young diagrams $(R_L,R_R)$.
Note that if we had a signature $3+1$ strings, we may have expected two directions for oscillation
modes, as there are two transverse directions to the strings.  However, in the $(2,2)$ signature
case, the transverse space to the worldsheet will have signature $(1,1)$ and it is natural to expect the string theory to have a symmetry to allow us to gauge away oscillation states in the transverse temporal direction, as we demand for this theory.
  We thus have the string states labeled by
$$(w,\theta, R_L,R_R)$$
where on shell states will be identified with the subset where $R_L=R_R=\emptyset$, which we simply
denote by $(w,\theta)$.  The corresponding field we will denote by $\Phi (w,\theta)$.  
This theory will have a conservation law associated to both $w$ and $\theta$.  
In particular all interactions conserve $\sum w_i$ as well as $\sum \theta_i$.  Moreover
the $\theta_i$ have an overalll 2-parameter redundancy by shifting the base point
on ${\bf T}^2$ where they are measured.  
This is then our proposal for computing the scattering amplitudes of this string theory:
Let $\langle \prod_{i=1}^N \Phi (w_i,\theta_i) \rangle $ denote the scattering amplitudes of this string.  Then we propose
$$\langle \prod_{i=1}^N \Phi (w_i,\theta_i) \rangle =I (w_i,\theta_i; q_1,q_2)$$
where $I$ denotes the superconformal index of the corresponding 5d theory.

The basic interaction of this theory for string states which have
primitive windings is cubic and is given by the refined topological vertex\footnote{The overall normalization of
the topological vertex, comes with some power of refined MacMahon function.  As explained
in \cite{iv} the natural normalization turns out to be associating a
 MacMahon function for each vertex.  Moreover each external string state comes
with an extra factor of inverse of MacMahon and the vacuum amplitude itself
come with a factor of square of MacMahon.  This gives an overall factor
of MacMahon to the power of twice the genus of the curve.}.  By an $SL(2,Z)$
transformation this will have the structure of\footnote{We are suppressing subtleties
having to do with choice of preferred directions \cite{ikv} for the refined case, which is a choice of gauge for off-shell scattering amplitudes.  See in particular \cite{onek}.}
$$\langle \Phi_{R^1_L,R^1_R}((1,0),\theta_1) \Phi_{R^2_L,R^2_R}((0,1),\theta_2) \Phi_{R^3_L,R^3_R}((-1,-1),\theta_3)\rangle =\delta(\sum \theta_i)\  C_{R^1_LR^2_LR^3_L}\
{\overline C}_{R^1_RR^2_RR^3_R}$$
This structure suggests that we should think of $\theta$ as some `momenta'
which is conserved (modulo integer multiples of $2\pi$).\footnote{Even though $\theta$
appears as position of the string in ${\bf T}^2$, its parameterization in terms of
the boson living on it leads to the identification $\partial \phi= i\theta$ which is the momentum for $\phi$.}

\subsection{Multi-wound strings}

In the context of toric geometries and the $(p,q)$ 5-brane web, the winding
strings are all primitive classes of $H_1({\bf T}^2,{\bf Z})$.    Moreover if
we consider a state in winding sector $n$ times a primitive one, we can
split it up to $n$ primitive strings.
However, it is natural
to ask if it makes sense to have `bound states'  were winding strings $w_i$ could end up
being multi-wound states.  From the view point of $(p,q)$ 5-branes at first
it seems not possible, because type IIB S-duality predicts that there are no bound states when
$p$ and $q$ are not relatively prime.  However, as we have already reviewed, introduction of 7-branes to the story,
can allow the 5-branes to effectively bind when a number of them are forced to end on a given
7-brane.   This has been considered in the context of $(p,q)$ 5-branes \cite{tach}.
The same story can be described in the geometric Calabi-Yau case:  There are
cases such as Calabi-Yau with higher del Pezzo's (such as ${\bf P}^2$ blown up at 8 points),
which have no toric description, but nevertheless they admit a degenerate torus action
as has been studied in \cite{druled}.  Having 7-branes suggest changing the
${\bf T}^2$ fiber, as in F-theory, to vary over the base.  Even though it is in principle
possible to consider this here, a more natural possibility is to take these 7-branes to be
infinitely far away in the ${\bf R}^2$ base of $M$, as we have considered
in our discussion of the 5-brane webs.  In such a case we can still take
the ${\bf T}^2 $ to have a fixed complex structure and just allow the boundary winding strings to be multi-wound.
Needless to say we can also consider the superconformal index for such theories
and use this to define the scattering of multi-wound strings.  In order to apply topological
vertex formalism, we need to extend it to include additional vertices.  To this end
additional vertex structure were introduced in \cite{druled}, precisely for computations
of topological string in these cases. Putting the wound strings as originating from
infinity suggests a somewhat different realization of the vertex of \cite{druled}.
In particular the generalized s-rule \cite{hw} studied in \cite{tach} would suggest that
we need two additional type of vertices involving multi-wound states.  One cubic
and one quartic.  See Figure \ref{etathree} for an example
of all three types of vertices.  The cubic vertex is of the form
$$(n,0)+(0,n)\rightarrow (n,n)$$
which reduces to the usual vertex when $n=1$, 
and the quartic one corresponds to
$$(n+m,0)+(0,n)\rightarrow (n,n)+(m,0)$$
and reduces to the cubic one when $m=0$.
It would 
be important to extend the topological vertex formalism to include these vertices
as well \cite{div}.

\subsection{Some basic properties of the scattering amplitudes}

As already discussed in section 3,  $\theta$-momenta are conserved.  Thus we end up with integration
over $\theta$ variables, one for each primitive loop of the scattering diagram, i.e. for each
genus of the worldsheet.    Moreover the propagator for each edge is fixed by
the external momenta, as well as the momenta in each loop.   The internal propagators
will include off-shell states associated to string oscillator modes.  In cases where
the external winding states are primitive, conservation
of winding number and the basic cubic interaction, fixes the genus
of the world sheet to be
$$g={1\over 2} \sum_{i<j} \langle w_i,w_j\rangle -{N\over 2} +1$$
where we order the $w_i$ according to the angle they make with the positive real axis in the counter-clockwise sense.  This formula is valid even if we have more than one string in the same
primitive winding class.
  If the external states include multiple wound strings, again
there is a fixed genus which gives the amplitudes but in this case the formula for $g$ is more complicated.  The above expression for $g$ gives an upper bound in the multi-wound case.

To compare with what we usually see in string theory, typically amplitudes
receive contributions for all genera.   Here, for each fixed scattering state, we get a contribution from a fixed genus.
This sometimes does happen for special supersymmetry protected scatterings in superstring
perturbation theory.  Here we are computing amplitudes with given fixed asymptotic
geometry, specified by $(w_i,\theta_i)$ and the 1-brane worldsheet is a Riemann surface which is forced to be a holomorphic brane (as in topological string).  This fixes the worldsheet to have a fixed genus.
Moreover the moduli of string worldsheet with this asymptotic behavior is not unique and we have
a $g$ complex dimensional moduli space of such world sheets (given by the classical
SW curve $\Sigma$).
Somehow, consistent with the fact that we do not want excitations in the temporal directions,
we should be forced to restrict
this to a $g$-real dimensional subspace. 

Another contrast with the usual string theory perturbation is that there we have an integration over
$3g-3$ complex moduli, which correspond to complexified lengths of the tubes making
up the worldsheet.  Here we are ending up with a $g$-dimensional real integration space,
as in field theories.  The integration variables
are comprised of the angular moduli of the tubes.  Moreover the angular moduli
of the edges are not independent and there is one such angular moduli for each loop (i.e. one
for each genus).

In string theory we can also have non-wound strings.
Here we do not yet have any interpretation for amplitudes associated to such states.  In that sense our worldsheet
is more like a brane which is embedded in the target space, rather than a map
from the 2d worldsheet to the target, as in Polyakov's string.

Note that for a given scattering theory, the worldsheet can in general be resolved to inequivalent
$(p,q)$ 5-branes, related by geometric transitions.  Clearly these theories lead
to the same conformal theory in 5d, because that corresponds to the same point
in the moduli space.  This in particular implies that if we compute the superconformal
index by different resolutions the answers will all be the same \cite{iv}.
This is of course a manifest symmetry of the string formulation as different resolutions
correspond to the same genus Riemann surface which are continuously deformable
to one another, without going through any singularities.

Finally, as we have explained in section 3, the target space $M$ is non-commutative
given by the bi-vector inverse to the $B$-field on $M$, which is proportional
to the imaginary part of the holomorphic (2,0) form on $M$ and our string can
be identified with a $(B,A,A)$ brane propagating in the background
of canonical coisotropic brane filling $M$.  So the needed ingredient is to make the
$(B,A,A)$ string dynamical.  This is similar to saying we have
a CFT and we are trying to make a string theory out of it.  We know
what the relevant CFT is.  The analog of symmetries and the ghost structure
of the string theory is what is missing.
Whatever this symmetry is, it should explain in particular that the integration over the $(B,A,A)$ brane
moduli and oscillators
are restricted to half the allowed dimensional space given by fixed asymptotics of the brane.

\subsection{Relation to Gaiotto theories}
Gaiotto considered a family of $N=2$ supersymmetric theories in 4 dimensions, related to $n$ M5 branes wrapping a Riemann
surface with punctures \cite{gaiotto}.  It was later noted that these theories do descend from 5d by choosing
suitable web of 5-branes \cite{tach}, or equivalently suitable CY geometries.  In this context
the partition function of the Gaiotto theories on $S^4$ is made of chiral blocks which
correspond to topological
string partition function on the corresponding Calabi-Yau.  In the 4d limit the chiral blocks reduce to
B-model topological string amplitudes studied in \cite{toda,mir}.  The chiral blocks of the
parent 5d theories which Gaiotto's
theories descend from has also been studied from the viewpoint of A-model topological
strings in \cite{wk}.

\begin{figure}[ptb]
\begin{center}
\includegraphics[
height=4.3215in,
width=4.1361in
]{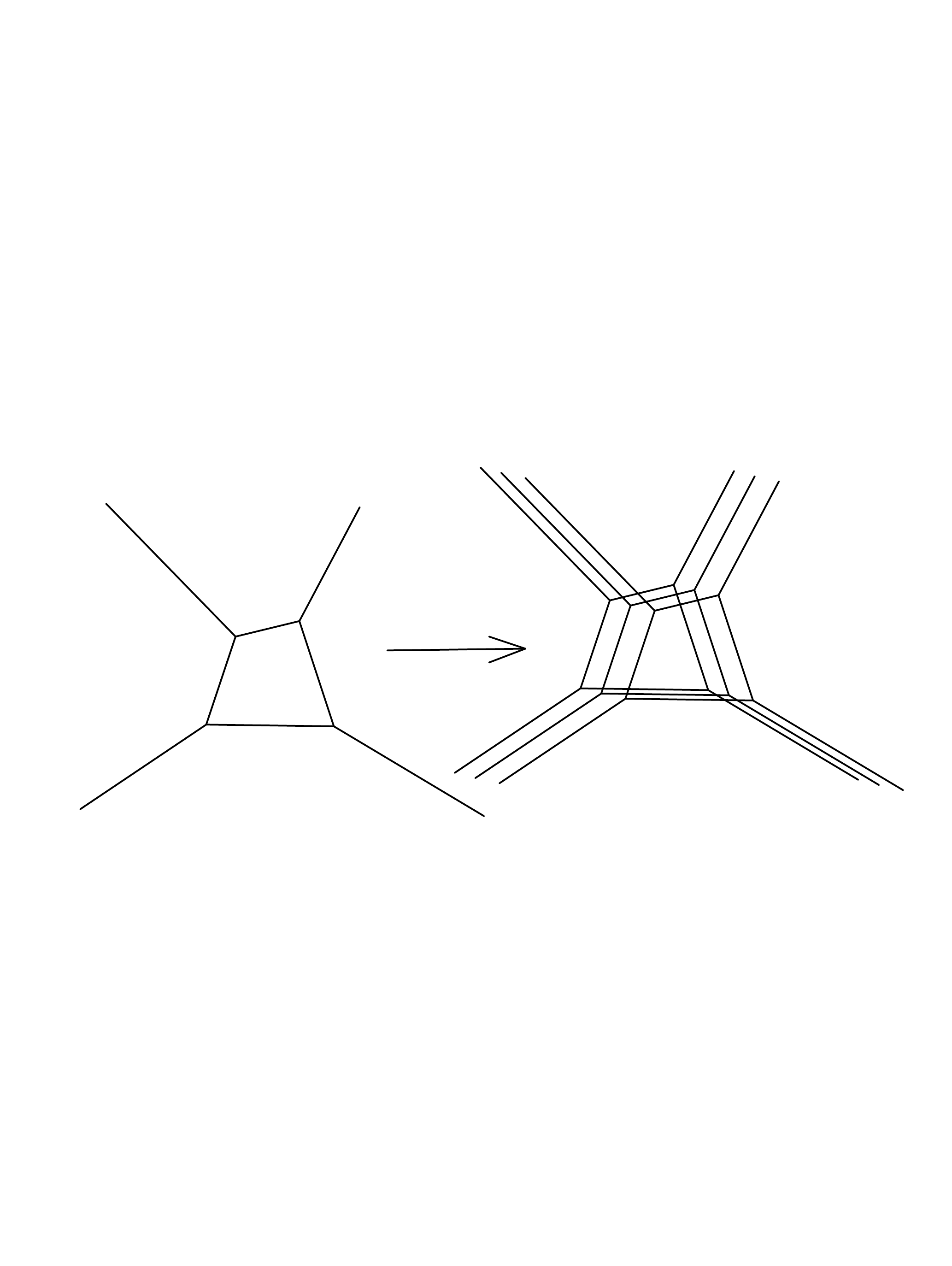}
\end{center}
\caption{{\protect\footnotesize {
Each web with primitive windings can be promoted to a
Gaiotto type theory, by changing the multiplicity by a factor of $n$.  
In this figure we are depicting the $n=3$ theory on a genus 1 curve with four punctures.
}}}%
\label{fig14}%
\end{figure}

The basic connection is very simple:  Consider any web diagram, and suppose
all the external lines are primitive winding states.  The corresponding curve can be viewed
as the Gaiotto curve\footnote{Note that the easiest curves to obtain will have at least 3 punctures
(corresponding to external lines.  However, it is also possible
to get rid of some or all the external lines, by identifying the ${\bf R}^2$ by a lattice
of shifts, making it effectively a ${\bf T}^2$ geometry \cite{akmv,hoiv}}.  Now change all the winding modes by a factor of $n$ in the  external states.  Clearly
this is still an allowed configuration of strings (see Figure \ref{fig14}).  However we still have to
decide for each external state with winding $w$, which winding state in $nw$ sector to choose. 
We may choose to split them to $n$ separate ones, i.e., $n$ strings in winding number
$1\cdot w$ which we denote
by $(1^n)$.   This corresponds to the `full' puncture of Gaoitto.
For example if we consider the three string states given by
windings $n[(1,0),(0,1),(-1,-1)]$ we get the 5d version of the $T_n$ theory.

More generally, we can choose to decompose the winding string to ($1^{k_1},..,n^{k_n}$) winding 
state, where
$$\sum_j  j k_j=n$$
This can be viewed as a partition of $n$ and fixes the corresponding Gaiotto puncture.
This is in accord with the web description of such punctures proposed in \cite{tach}.
Note that the symmetries of this sector is manifestly $U(k_1)\times U(k_2)\times... \times U(k_n)$
corresponding to the fact that the mass parameters live in the Cartan torus of this group
due to symmetries of exchanging the strings with the same winding number.   This structure
is reminiscent of matrix strings \cite{dmat}
where conjugacy classes of $S_n$ permutation group labels the symmetric product of string
states, which decompose according to length of the cycle in each conjugacy class.
To actually obtain the 4d limit we need to consider the case where the masses (or the
corresponding $\theta_i$ in the 5d) come together.  The $U(1)$ center of mass of the
strings decouples and we are left with the relative separation of the strings
as the parameters. 

From the perspective of the present paper choosing the $n$ multiplicity for
all winding states is very special.  Clearly in 5d there are additional conformal
theories corresponding to different winding multiplicities for different external
states and it is natural to expect that a large class of such theories yield new additional
$N=2$ superconformal theories in 4d in similar limits.    It would be interesting
to study this question and see how many of these additional 5d CFT's
lead to new 4d CFT's.   Needless to say dualities, similar to the ones
studied by Gaiotto will hold for this class of 4d theories as well, which is manifest
from the Riemann surface (or web) description of it.

Surface operators in the 4d can also be introduced from the perspective of this string
theory.  Namely we introduce the corresponding spectator branes which inject
impurities on the world sheet of strings (corresponding to fermionic operators)
as already discussed.  This modifies the scattering amplitudes of the strings depending
on which impurities we have inserted.  This can be viewed as the computation of the
superconformal index in 5 dimensions in the presences of codimension 2 defects,
as discussed in \cite{iv}.

\subsection{Connections with AGT Conjecture}

According to the AGT conjecture the partition function of $A_1$ Gaiotto theory on $S^4$ reduces to Liouville
amplitudes on the Gaiotto curve \cite{agt}.  Similarly the partition function of $A_n$ theories
are conjectured to correspond to Toda partition functions \cite{wyl}.  It is natural
to ask how these fit with the scattering amplitudes of our string theory.  After
all, as mentioned in the previous section, for special scattering amplitudes
in particular limits they should reduce to the corresponding partition function.
Here we review the connection between these two setups along the lines suggested in
\cite{toda}.

For simplicity let us consider the $A_1$ case, and in the limit $q_1=q_2^{-1}$, i.e.
turning off the refinement of topological string.  In this situation, we consider a web
configuration which we identify with the analog of Gaiotto curve, and double the
number of strings associated with each edge.  As discussed before this would
correspond to the $A_1$ theory.  Let us focus along one of the edges of this string.
We thus have two parallel tubes which have coalesced.  This is reminiscent of something
else we have already discussed:  the defect operators.  If we separated the two parallel tubes
and take one infinitely far away, and if we introduced D3 branes suspended between the
tubes, this corresponds to insertion of fermionic operator on the tube.  Taking
into account that we have both the topological string and the anti-topological string,
this would mean the insertion of $\psi_L \psi_R={\rm exp}(i\phi)$.   Now recall that
the spectator brane is not really a spectator and is part of the web.  Thus the D3 branes
ending on the spectator brane introduce insertion of fermions also on the spectator
brane (but complex conjugate
relative to the one on the first brane).   Thus we expect insertion of operators of the type

\begin{figure}[ptb]
\begin{center}
\includegraphics[
height=4.3215in,
width=4.1361in
]{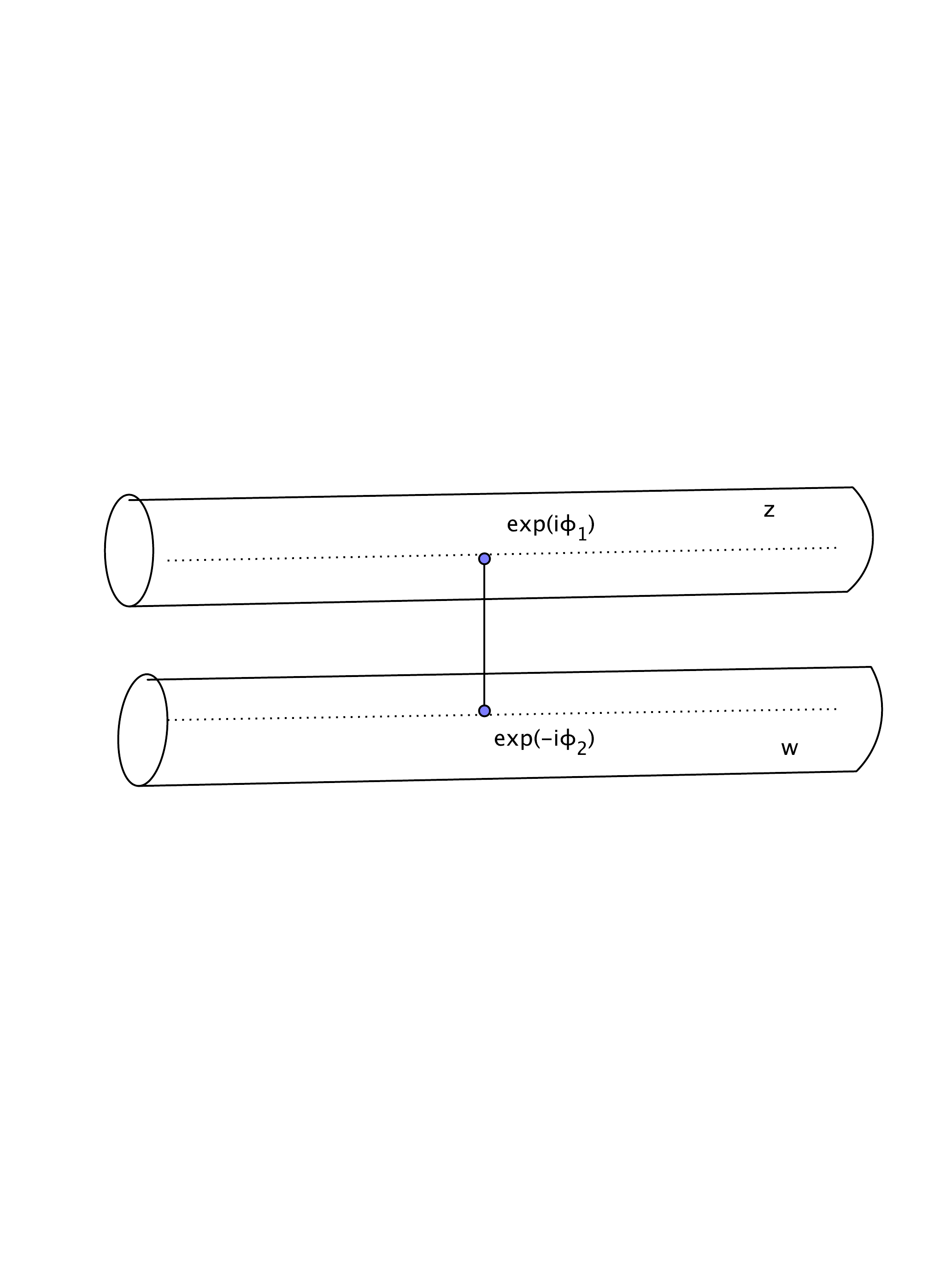}
\end{center}
\caption{{\protect\footnotesize {When two parallel 5-branes come close
there are instanton corrections corresponding to insertion of
fermion operators on either end.  The instanton can slide on the tubes.
The Liouville potential arises
when the tubes coincide.}}}%
\label{slide}%
\end{figure}
$$\psi^1_L\psi^1_R(z,{\overline z}) \psi^{2*}_L\psi^{2*}_R(w,{\overline w})={\rm exp}[i(\phi_1(z)]\ {\rm exp}[-i\phi_2(w)]$$
Moreover if we bring the tubes back together and make them coalesce the D3 branes
that we previously added by hand have zero tension and can be viewed as an instanton
contribution to this background (see Figure \ref{slide}).   Moreover since
the tubes are parallel, the D3 brane can slide along the entire tube and thus the
surface operator could be along the entire Gaiotto tube $z=w$.  This simply means that we add
the term
$$\int_{worldsheet} d^2z \ {\rm exp}[i(\phi_1-\phi_2)]$$
to the action.
Adding this term to the action simply ensures that we can have an arbitrary number of
the `tensionless' D3 branes suspended between the two tubes.   Moreover putting
a parameter $\mu$ in front of this term in the action, counts how may of them we bring
down, which should be related to the separation between them $m$, i.e.  the length
of the D3 brane segment.   In other words $\mu ={\rm exp} (-m)$.   Rescaling $\mu$ is thus
equivalent to rescaling the separation between the lines, which is the same as
changing the momentum of Liouville insertions.  This is the same as what is familiar
in Liouville theory:  That $\mu$ can be absorbed to the choice of external momenta.
Here we have talked about the unrefined case.  As discussed in \cite{toda} going
to the refined case in the 4d limit would end up introducing the background charge for the $\phi$ and
changing the coefficient of the exponent in such a way that it corresponds to a $(1,1)$ operator.
Similarly this can be carried over to the Toda case, where we get insertion of
multi scalars, one for each tube, leading up to the potential term for the Toda.\footnote{Note that here we have the complexified version of Liouville \cite{comp}, as the exponents
involve complex parameters. This is related to the fact that the coupling
constants for us correspond to pure rotations and are purely imaginary.  Analytically continuing
this to the make them real is what corresponds to the radius of compactification in the
$S^4$ partition function.}

Note that the integration over the internal momenta of the Liouville theory (or Toda theory)
map to the integral over the moduli space of the Riemann surface which is the appropriate
cover of the Gaiotto curve.  In our set up this integral is simply all the allowed worldsheets
compatible with a given asymptotic string scattering states.

\begin{figure}[ptb]
\begin{center}
\includegraphics[
height=4.3215in,
width=4.1361in
]{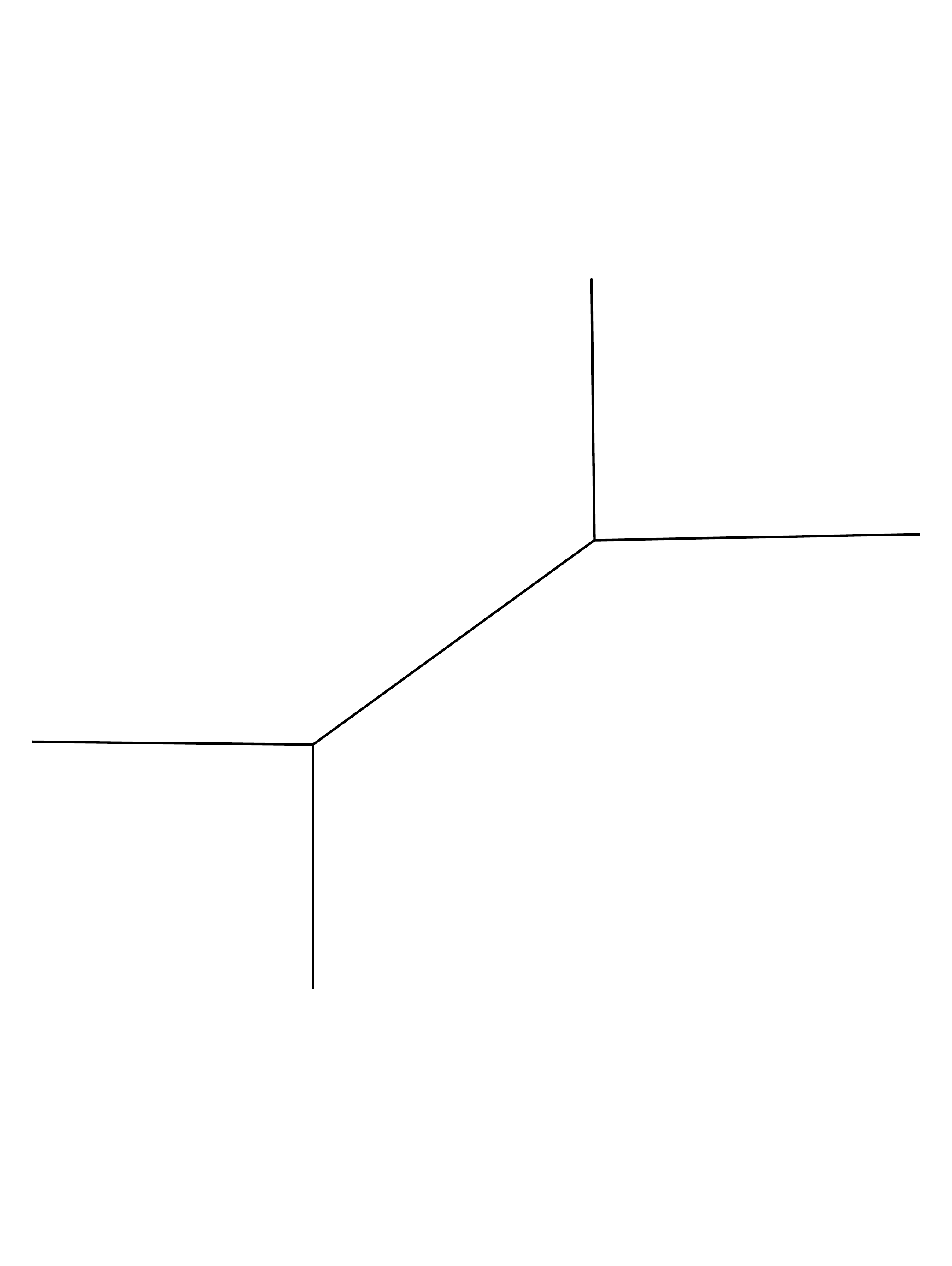}
\end{center}
\caption{{\protect\footnotesize {
A simple scattering amplitude with no internal loops.
}}}%
\label{example1}%
\end{figure}

\subsection{Examples}

Here we consider a few simple examples of scattering amplitudes.
The simplest type of 2-string scatterings corresponds to the following amplitude (see Figure
\ref{example1}):
$$(1,0)+(0,1)\rightarrow (1,0)+(0,1)$$
which can be viewed as the amplitude for
$${\cal A} =\langle \Phi((1,0),\theta_1)\Phi((0,1),\theta_2)\Phi((-1,0),\theta_3)\Phi((0,-1),\theta_4)\rangle$$
Using momentum conservation, and shifting two of the $\theta_2=\theta_3=0$, the amplitude
will only depend on $\theta_1=\theta$.   Dropping the delta function, we have the following
answer of this scattering:
\begin{figure}[ptb]
\begin{center}
\includegraphics[
height=4.3215in,
width=4.1361in
]{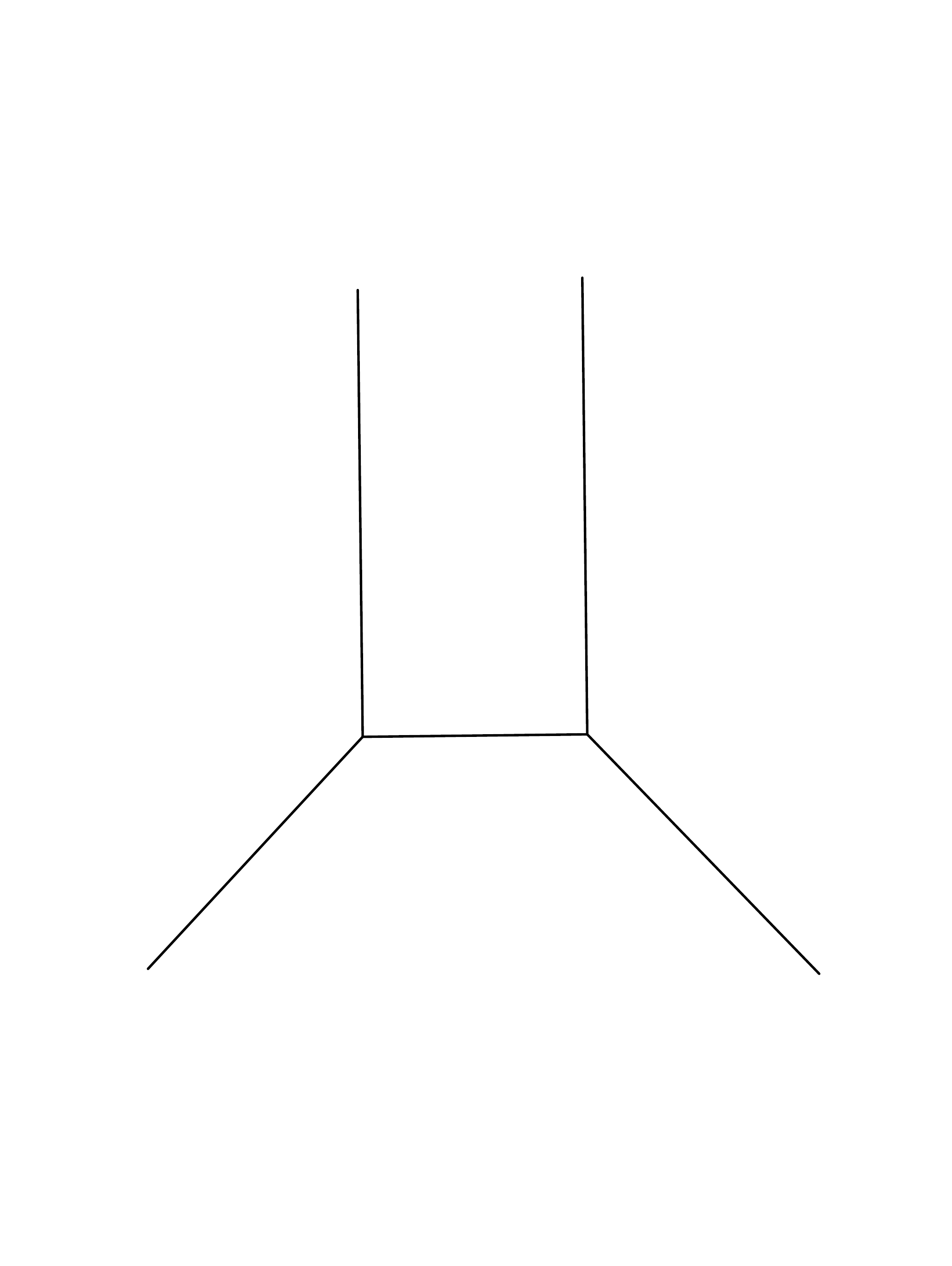}
\end{center}
\caption{{\protect\footnotesize {
Another simple scattering amplitude with no internal loops.
}}}%
\label{example2}%
\end{figure}

$${\cal A}=\prod_{i,j,k,l=1}^{\infty} (1-q_1^iq_2^j \ e^{i\theta})(1-q_1^k q_2^l\ e^{-i\theta})$$
This corresponds to the $O(-1)+O(-1)$ geometry over ${\bf P}^1$ and has $g=0$.  Another simple two
particle $g=0$ scattering is (see Figure \ref{example2})
$$(1,0)+(1,0)\rightarrow (1,-1)+(1,1)$$
leading to the amplitude
$${\cal A}=\prod_{i,j,k,l=1}^{\infty} {1\over (1-q_1^iq_2^j \ e^{i\theta})(1-q_1^k q_2^l\ e^{-i\theta})}$$

\begin{figure}[ptb]
\begin{center}
\includegraphics[
height=4.3215in,
width=4.1361in
]{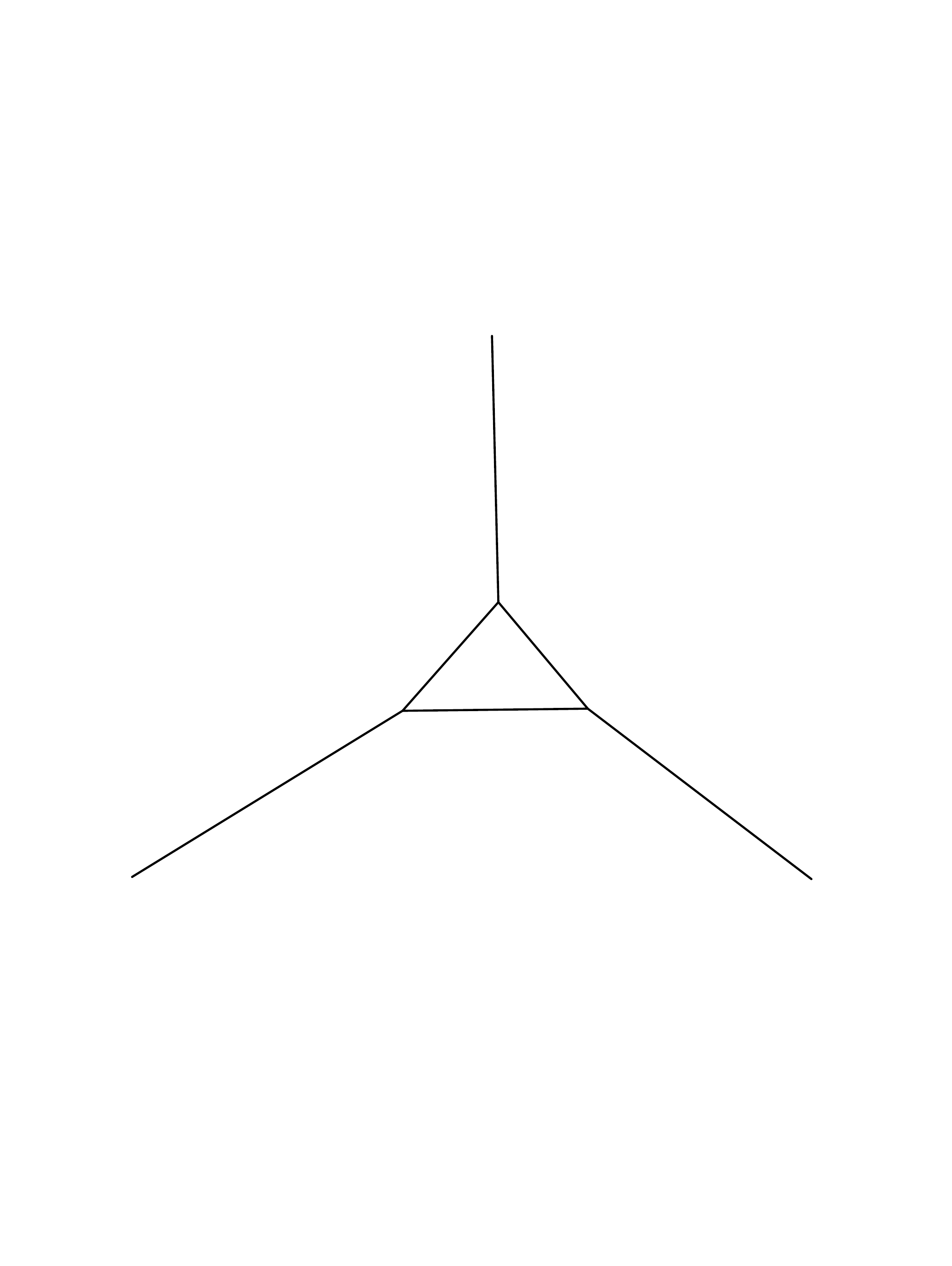}
\end{center}
\caption{{\protect\footnotesize {
The decay amplitude with one internal loop.  The amplitude depends on no external
momenta.
}}}%
\label{example3}%
\end{figure}

The higher genus amplitudes are more difficult to compute, though they can be computed.
The simplest  scattering corresponding to genus 1 corresponds to a three particle
amplitude (see Figure \ref{example3}):
$$(-2,1)\rightarrow (-1,2)+(-1,-1)$$
This scattering amplitude has no mass parameter $(N-3=3-3=0)$ so the
amplitude depends only on $q_1,q_2$.  It corresponds to the superconformal
index of M-theory on ${\bf C}^3/Z_3$.  It can be computed using refined topological
string (see \cite{iv}).

One can also compute amplitudes for non-toric geometries, for example
corresponding to del Pezzo with 8 points blown up with $E_8$ global symmetry
(corresponding to $SU(2)$ with 7 flavors \cite{sei,dkv,mseib}.
Using the result of \cite{tach} this can be viewed as the result of the scattering
of non-primitive winding modes:
$$(3,0)+(3,0)+(0,2)+(0,2)+(0,2)\rightarrow (1,1)+(1,1)+(1,1)+(1,1)+(1,1)+(1,1)$$
This has 8 parameters (11-3), and they can be identified with the Cartan
of $E_8$ (or 7 masses of the fundamental $SU(2)$'s and one coupling constant
of  $SU(2)$) and has a genus $1$ worldsheet.  This scattering amplitude has been recently computed using localization techniques for gauge theory amplitudes in \cite{kimyeong}.

\section{Open Questions}

The primary question is to give a direct definition of the string theory we have
provided evidence for.  It is also natural to ask what is the relation of this string
theory to $N=2$ string theory \cite{ov}?\footnote{In this context it is useful
to recall that $N=2$ string theory is equivalent to $N=4$ topological string theory \cite{berkv}.
In fact using this formalism the
partition function of $N=2$ string has been computed in the background of
$T^*{\bf T}^2$ \cite{allloop}.}.  The fact that both are defined in 4 dimensions
with a complex structure suggests some relation between the two.  In fact there are more
common roots:  They both seem to be related to integrable structures.  Self-dual geometries
and connections, which is the backbone of $N=2$ string theory is deeply rooted in integrable
structures. On the other hand the  worldsheet which underlies the scattering
amplitude of the string we have proposed, is related to quantum version of Seiberg-Witten
curve which again is deeply connected with integrable structures (see in particular
 \cite{nekshat}).
 Both the $N=2$ strings
and the string being proposed here have effectively the same on-shell degrees of freedom corresponding
to one scalar field in 4d.  This is because choosing a winding vector has effectively
two degrees of freedom (corresponding to integral points on ${\bf R}^2$) and in
addition the momentum $\theta$, leading to three parameters to vary, just
as one would expect for a massless scalar field in 4d which is the on shell
degree of freedom of $N=2$ strings.
It could be that to connect the two we need to introduce additional background branes
in the $N=2$ string, as in the space-time filling coisotropic brane, making the space-time
effectively non-commutative.

It is also clear that some aspects of the vertex for the string theory we have
proposed need to be developed further.  In particular currently we do not have a
simple vertex which capture multiple winding strings.  In particular
as we have noted we need to develop techniques to compute an additional
cubic and quartic type vertices.  It would be important to
extend the topological vertex formalism to this case.  Also aspects of the refined topological
string need to be better understood.  In particular one needs to extend the non-commutative background
to include both coupling constants in the refined case. 

It is also natural to consider more general target space.  For example
it would be interesting to study this when $M={\bf C}^2$.  This
would correspond to SUSY amplitudes directly in 4d.   Also it would to be
interesting to consider $M$ being an arbitrary
hyperK\"ahler geometry.  The fact that we can introduce 7-branes, already
suggests that the elliptic fiber can vary (as the ${\bf T}^2$ is identified with the
F-theory fiber), which is already a hyperK\"ahler manifold which is not flat.
Even though for the purposes of our paper we were able to push the 7-branes to infinity and treated
the geometry as flat, this already shows that there is no inconsistency
with having a non-flat hyperK\"ahler geometry.   Moreover the description
in terms of $(B,A,A)$ branes in the background of cc branes supports
this possibility, because these can be defined for arbitrary hyperK\"ahler manifolds.
Perhaps it would be natural to leave two directions
non-compact to correspond to the time directions.  Even this restriction would allow
some interesting additional possibilities:  For example what would this theory correspond to when
$M$ is the non-commutative version of $T^*\Sigma$ where $\Sigma$ is an arbitrary
Riemann surface?  Note that this space does admit a complete $(2,2)$ signature hyperK\"ahler metric
\cite{ov}, where the fiber directions correspond to time-like components of the metric.

It is also natural to ask if there is twistorial version of this theory.  Namely, we have
chosen a constant background B-field in 4-dimensional flat space.  Different
choices of such a $B$, where we can integrate over, can be viewed as a twistorial version of
this theory.\footnote{This is reminiscent of the recent work \cite{heckman}
which considers a family of non-commutative parameters.}

The results presented here suggest that studying supersymmetric partition functions
for a family of quantum theories in a given dimension may lead to a single unifying
theory.  In the 5d case, we found a string theory.   I conjecture
that this structure will also be true in other dimensions.  It would be interesting to study
this question and ask which unifying quantum theory, if any, underlies the supersymmetric partition functions for a family of quantum theories in a given dimension.

\section*{Acknowledgments}

I would like to thank the SCGP where this work was initiated in the 10-th
Simons Workshop on math and physics.  In addition to the talks presented there,  I have benefitted from discussions
with M. Aganagic, S. Cecotti, M. Cheng, E. Diaconescu, R. Dijkgraaf, B. Haghighat, J. Heckman, S. Hellerman, A. Iqbal, D. Jafferis, G. Lockhart, H. Ooguri, and V. Pestun.

This work is supported in part by NSF grant PHY-0244821. 

\bibliographystyle{utcaps}	
\bibliography{myrefs}	

\end{document}